\documentclass[review]{elsarticle}
\UseRawInputEncoding
\usepackage{lineno,hyperref}
\usepackage{graphicx}
\usepackage{amsmath,bm}
\usepackage{url}
\usepackage[hyphenbreaks]{breakurl}
\usepackage{array}
\usepackage{booktabs}
\journal{Neurocomputing}

\begin{document}

\begin{frontmatter}

\title{Skin disease diagnosis with deep learning: a review}

\author[mymainaddress]{Hongfeng Li\corref{mycorrespondingauthor}}
\cortext[mycorrespondingauthor]{Corresponding author}
\ead{lihongfeng@math.pku.edu.cn}

\author[mysecondaryaddress]{Yini Pan}

\author[mymainaddress]{Jie Zhao}

\author[mythirdaddress]{Li Zhang}

\address[mymainaddress]{Center for Data Science in Health and Medicine, Peking University, Beijing 100871, China}
\address[mysecondaryaddress]{Academy for Advanced Interdisciplinary Studies, Peking University, Beijing 100871, China}
\address[mythirdaddress]{Center for Data Science, Peking University, Beijing 100871, China}

\begin{abstract}
Skin cancer is one of the most threatening diseases worldwide. However, diagnosing skin cancer correctly is challenging. Recently, deep learning algorithms have emerged to achieve excellent performance on various tasks. Particularly, they have been applied to the skin disease diagnosis tasks. In this paper, we present a review on deep learning methods and their applications in skin disease diagnosis. We first present a brief introduction to skin diseases and image acquisition methods in dermatology, and list several publicly available skin datasets for training and testing algorithms. Then, we introduce the conception of deep learning and review popular deep learning architectures. Thereafter, popular deep learning frameworks facilitating the implementation of deep learning algorithms and performance evaluation metrics are presented. As an important part of this article, we then review the literature involving deep learning methods for skin disease diagnosis from several aspects according to the specific tasks. Additionally, we discuss the challenges faced in the area and suggest possible future research directions. The major purpose of this article is to provide a conceptual and systematically review of the recent works on skin disease diagnosis with deep learning. Given the popularity of deep learning, there remains great challenges in the area, as well as opportunities that we can explore in the future.
\end{abstract}


\end{frontmatter}

%\linenumbers

\section{Introduction}

%\paragraph{Installation} If the document class \emph{elsarticle} is not available on your computer, you can download and install the system package \emph{texlive-publishers} (Linux) or install the \LaTeX\ package \emph{elsarticle} using the package manager of your \TeX\ installation, which is typically \TeX\ Live or Mik\TeX.

Skin disease is one of the most common diseases among people worldwide. There are various types of skin diseases, such as basal cell carcinoma (BCC), melanoma, intraepithelial carcinoma, and squamous cell carcinoma (SCC)~\cite{gandhi2015skin}. Particularly, skin cancer has been the most common cancer in United States and researches showed that one-fifth of Americans will suffer from a skin cancer during their lifetime~\cite{guy2015vital,stern2010prevalence}. Melanoma is reported as the most fatal skin cancer with a mortality rate of $1.62\%$ among other skin cancers~\cite{tarver2012american}. According to the American Cancer Society's estimates for melanoma in the United States for 2020, there will be about $100,350$ new cases of melanoma and $6,850$ people are expected to die of melanoma~\cite{cancerq2020}. On the other hand, BCC is the most common skin cancer, and although not usually fatal, it places large burdens on health care services~\cite{lomas2012systematic}. Fortunately, early diagnosis and treatment of skin cancer can improve the five-year survival rate by around $14\%$~\cite{ali2012systematic}.

\begin{figure}
\center
\includegraphics[width=0.6\textwidth]{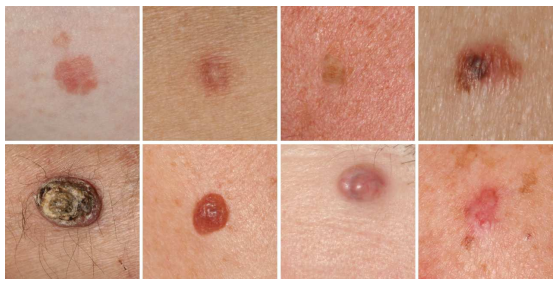}
\caption{Several examples of different types of skin diseases. These images come from the Dermofit Image Library~\cite{dermofit}.} \label{skin}
\end{figure}

However, diagnosing a skin disease correctly is challenging since a variety of visual clues, such as the individual lesional morphology, the body site distribution, color, scaling and arrangement of lesions, should be utilized to facilitate the diagnosis. When the individual components are analyzed separately, the diagnosis process can be complex~\cite{habif2017skin}. For instance, there are four major clinical diagnosis methods for melanoma: ABCD rules, pattern analysis, Menzies method and $7$-Point Checklist. Often only experienced physicians can achieve good diagnosis accuracy with these methods~\cite{whited1998does}. The histopathological examination on the biopsy sampled from a suspicious lesion is the gold standard for skin disease diagnosis. Several examples of different types of skin diseases are demonstrated in Fig.~\ref{skin}. Developing an effective method that can automatically discriminate skin cancer from non-cancer and differentiate skin cancer types would therefore be beneficial as an initial screening tool.

Differentiating a skin disease with dermoscopy images may be inaccurate or irreproducible since it depends on the experience of dermatologists. In practice, the diagnostic accuracy of melanoma from the dermoscopy images by an inexperienced specialist is between $75\%$ to $84\%$~\cite{ali2012systematic}.  One limitation of the diagnosis performed by human experts is that it heavily depends on subjective judgment and varies largely among different experts. By contrast, a computer aided diagnostic (CAD) system is more objective. By utilizing handcrafted features, traditional CAD systems for skin disease classification can achieve excellent performance in certain skin disease diagnosis tasks~\cite{arroyo2014automated,madooei2016incorporating,saez2014pattern}. However, these systems usually focus on limited types of skin diseases, such as melanoma and BCC. Therefore, they are typically unable to be generalized to perform diagnosis over broader classes of skin diseases. The reason is that the handcrafted features are not suitable for a universal skin disease diagnosis. On one hand, handcrafted features are usually specifically extracted for limited types of skin diseases. They can hardly be adapted to other types of skin diseases. One the other hand, due to the diversity of skin diseases, human-crafted features cannot be effective for every kind of skin disease~\cite{habif2017skin}. Feature learning can be one solution to this problem, which eliminates the need of feature engineering and extracts effective features automatically~\cite{bengio2013representation}. Many feature learning methods have been proposed in the past few years~\cite{chang2015stacked,cruz2013deep,arevalo2015unsupervised}. However, most of them were applied on dermoscopy or histopathology images processing tasks and mainly focused on the detection of mitosis and indicator of cancer~\cite{wang2014cascaded}.

Recently, deep learning methods have become popular in feature learning and achieved excellent performances in various tasks, including image classification~\cite{he2016deep,krizhevsky2012imagenet}, segmentation~\cite{badrinarayanan2017segnet,chen2017deeplab}, object detection~\cite{ouyang2015deepid,li2016deep} and localization~\cite{sermanet2013overfeat,ren2015faster}. A variety of researches [9, 23, 12, 27, 25] showed that the deep learning methods were able to surpass humans in many computer vision tasks. One thing behind the success of deep learning is its ability to learn semantic features automatically from large-scale datasets. In particular, there have been many works on applying deep learning methods to skin disease diagnosis~\cite{esteva2015deep,codella2017deep,esteva2017dermatologist,binder1994application,liao2016skin}. For example, Esteva et al.~\cite{esteva2015deep} proposed a universal skin disease classification system based on a pretrained convolutional neural network (CNN). The top-$1$ and top-$3$ classification accuracies they achieved were $60.0\%$ and $80.3\%$ respectively, which significantly outperformed the performances of human specialists. Deep neural networks can deal with the large variations included in the images of skin diseases through learning effective features with multiple layers. Despite these technological advances, however, lack of available huge volume of labeled clinical data has limited the wide application of deep learning in skin disease diagnosis.

In this paper, we present a comprehensive review of the recent works on deep learning for skin disease diagnosis. We first give a brief introduction to skin diseases. Through literature research, we then  introduce common data acquisition methods and list several commonly used and publicly available skin disease datasets for training and testing deep learning models. Thereafter, we describe the basic conception of deep learning and present the popular deep learning architectures. Accordingly, prevalent deep learning frameworks are described and compared. To make it clear that how to evaluate a deep learning method, we introduce the evaluation metrics according to different tasks. We then draw on the literature of applications of deep learning in skin disease diagnosis and introduce the content according to different tasks. Through analyzing the reviewed literature, we present the challenges remained in the area of skin disease diagnosis with deep learning and provide guidelines to deal with these challenges in the future. Considering the lack of in-depth comprehension of skin diseases and deep learning by broader communities, this paper could provide the understanding of the major concepts related to skin disease and deep learning at an appropriate level. It should be noted that the goal of the review is not to exhaust the literature in the field. Instead, we summarize the related representative works published before/in the year 2019 and provide suggestions to deal with current challenges faced in the field by referring recent works until the year 2020.

Compared with previous related works, the contributions in this paper can be summarized as follows. First, we systematically introduce the recent advances in skin disease diagnosis with deep learning from several aspects, including the skin disease and public datasets, concepts of deep learning and popular architectures, applications of deep learning in skin disease diagnosis tasks. Though there have been papers that reviewed works on skin disease diagnosis, some of them~\cite{pathan2018techniques} focused on traditional machine learning and deep learning only occupied a small section of them. Alternatively, others~\cite{brinker2018skin} only discussed specific skin diseases diagnosis task (e.g., classification) and the presented deep learning methods were out of date. By contrast, this paper provides a systematic survey of the field of skin disease diagnosis focusing on recent applications of deep learning. With this article, one could obtain an intuitive understanding of the essential concepts of the field of skin disease diagnosis with deep learning. Second, we present discussions about the challenges faced in the field and suggest several possible directions to deal with these issues. These can be taken into consideration by ones who are willing to work further in this field in the future.

The remainder of the paper is structured as follows. Section $2$ briefly introduces the skin disease and Section $3$ touches upon the common skin image acquisition methods and available public skin disease datasets for training and testing deep learning models. In section $4$, we introduce the conception of deep learning and popular architectures. Section $5$ briefly introduces the common deep learning frameworks and evaluation metrics for testing the effectiveness of an algorithm are presented in section $6$. After that, we investigate the applications of deep learning methods in skin disease diagnosis according to the types of tasks in section $7$. Then we highlight the challenges in the area of skin disease diagnosis with deep learning and suggest future directions dealing with these challenges in section $8$. Finally, we conclude the article in Section $9$.

\section{Skin disease}

Skin is the largest immense organ of the human body, consisting of epidermis, dermis and hypodermis. The skin has three main functions: auspice, sensation and thermoregulation, providing an excellent aegis against aggression of the environment. Stratum corneum is the top layer of the epidermis and optically neutral protective layer with varying thickness. The stratum corneum consists of keratinocytes that produce keratin responsible for benefiting the skin to protect the body. The incident of light on the skin is scattered due to the stratum corneum. The epidermis includes melanocytes in its basal layer. Particularly, melanocytes make the skin generate pigment called as melanin, which provides the tan or brown color of the skin. Melanocytes act as a filter and protect the skin from harmful ultraviolet (UV) sunrays by generating more melanin. The extent of absorption of UV rays depends on the concentration of melanocytes. However, the unusual growth of melanocytes causes melanoma. The dermis is located at the middle layer of the skin, consisting of collagen fibers, sensors, receptors, blood vessels and nerve ends. It provides elasticity and vigor to the skin~\cite{pathan2018techniques}.

Deoxyribonucleic acid (DNA) consists of molecules called nucleotides. A nucleotide comprises of a phosphate and a sugar group along with a nitrogen base. The order of nitrogen bases in the DNA sequence forms the genes. Genes decide the formation, multiplication, division and death of cells. Oncogenes are responsible for the multiplication and division of cells. Protective genes are known as tumor suppressor genes. Usually, they inhibit cell growth by monitoring how expeditiously cells divide into incipient cells, rehabilitating mismatched DNA and controlling when a cell dies. The uncontrollability of a cell occurs due to the mutation of the tumor suppressor genes, eventually forming a mass called tumor (cancer). UV rays can damage the DNA, which causes the melanocytes to produce melanin at a high abnormal rate. Appropriate amount of UV rays benefits the skin to form vitamin D, but excess will cause pigmented skin lesions~\cite{uong2010melanocytes}. Particularly, the malignant tumor occurred due to abnormal growth of the melanocytes is called as melanoma~\cite{feng2013studies}.

There are three major types of skin cancers, i.e., malignant melanoma (MM), squamous cell carcinoma, and basal cell carcinoma. In particular, the latter two are developed from basal and squamous keratinocytes and also known as keratinocyte carcinoma (KC). They are the most commonly occurring skin cancers in men and women, with over $4.3$ million cases of BCC and $1$ million cases of SCC diagnosed each year in the United States, although these numbers are likely to be underestimated~\cite{rogers2015incidence}. However, MM, an aggressive malignancy of melanocytes, is a less common but far more deadly skin cancer. It often starts as minuscule, with a gradual change in size and color. The color of melanin essentially depends on its localization in the skin. The color ebony is due to melanin located in the stratum corneum. Light to dark brown, gray to gray-blue and steel-blue are observed in the upper epidermis, papillary dermis and reticular dermis respectively. In case of  benign lesions, the exorbitant melanin deposit presents in the epidermis. Melanin presence in the dermis is the most consequential designation of melanoma causing prominent vicissitude in skin coloration. There are several other designations for melanoma, including thickened collagen fibers in addition to pale lesion areas with a large blood supply at the periphery. The gross morphologic features additionally include shape, size, coloration, border and symmetry of the pigmented lesion. Biopsy and histology are required to perform explicit diagnosis in case the ocular approximation corroborates a suspicion of skin cancer~\cite{gomez2007independent}. According to microscopic characterizations of the lesion, there are four major categories of melanoma, i.e., superficial spreading melanoma (SSM), nodular melanoma (NM), lentigo malignant melanoma (LMM) and acral lentiginous melanoma (ALM).

\section{Image acquisition and datasets}

\subsection{Image acquisition}

Dermatology is termed as a visual specialty wherein most diagnosis can be performed by visual inspection of the skin. Equipment-aided visual inspection is important for dermatologists since it can provide crucial information for precise early diagnosis of skin diseases. Subtle features of skin diseases need further magnification such that experienced dermatologists can visualize them clearly~\cite{marghoob2012atlas}. In some cases, a skin biopsy is needed which provides the opportunity for a microscopic visual examination of the lesion in question. Lots of image acquisition approaches were developed to facilitate dermatologists to overcome problems caused by apperception of minuscule sized skin lesions.

Dermoscopy, one of the most widely used image acquisition methods in dermatology, is a non-invasive imaging technique that allows the visualization of skin surface by the light magnifying device and immersion fluid~\cite{pellacani2002comparison}. Statistics shows that dermoscopy has improved the diagnosis performance of malignant cases by $50\%$~\cite{sinz2017accuracy}. Kolhaus was the first one to start skin surface microscopy in 1663 to inspect minuscule vessels in nail folds~\cite{stolz2002color}. The term dermatoscopy was coined by Johann Saphier, a German dermatologist, in 1920 and then dermatoscopy is employed for skin lesion evaluation~\cite{sacchidanand2013nail}. Dermoscopy additonally kenned as epiluminescence microscopy (ELM) is a non-invasive method that can be utilized in vivo evaluation of colors and microstructure of the epidermis. The dermo-epidermal junction and papillary dermis cannot be observed by unclad ocular techniques~\cite{soyer2011dermoscopy}. These structures form the histopathological features that determine the level of malignancy and indicate whether the lesion is necessary to be biopsied~\cite{noor2009dermoscopy}. The basic principal of dermoscopy is transillumination of the skin lesion. The stratum corneum is optically neutral. Due to the incidence of visible radiation on the surface of skin, reflection occurs at the stratum corneum air interface~\cite{jerant2000early}. Oily skin enables light to pass through it; therefore, linkage fluids applied on the surface of the skin make it possible to magnify the skin and access to deeper layers of the skin structures~\cite{erdei2010new}. However, the scope of observable structures is restricted compared with other techniques, presenting a potentially subjective diagnosis precision. It was shown that the diagnosis precision depended on the experience of dermatologists~\cite{lorentzen1999clinical}. Dermoscopy is utilized by most of the dermatologists in order to reduce patient concern and present early diagnosis.

In vivo, the confocal laser scanning microscopy (CLSM), a novel image acquisition equipment, enables the study of skin morphology in legitimate period at a resolution equal to that of the traditional microscopes~\cite{gerger2005diagnostic}. In CSLM, a focused laser beam is used to enlighten a solid point inside the skin and the reflection of light starting there is measured. Gray-scale image is obtained by examining the territory paralleling to the skin surface. According to the review~\cite{guitera2012vivo}, a sensitivity of $88\%$ and specificity of $71\%$ were obtained with CSLM. However, the confocal magnifying lens in CSLM involves high cost (up to $\$50,000$ to $\$100,000$).

Optical coherence tomography (OCT) is a high-determination non-obtrusive imaging approach that has been utilized in restorative examinations. The sensitivity and specificity vary between $79\%$ to $94\%$ and $85\%$ to $96\%$, respectively~\cite{fujimoto2003optical}. The diagnosis performed with OCT is less precise than that of clinical diagnosis. However, a higher precision can be obtained for distinguishing lesions from the normal skin.

The utilization of a skin imaging contrivance is referred as spectrophotometric or spectral intracutaneous analysis (SIA) of skin lesions. The SIA scope can improve the performance of practicing clinicians in the early diagnosis of the deadly disease. A study has reported that SIA scope presented the same sensitivity and specificity as these of dermatoscopy performed by skilled dermatologists~\cite{blum2004value}. The interpretation of these images is laborious due to the involution of the optical processes involved.

Ultrasound imaging~\cite{passmann1996100} is an important tool for skin disease diagnosis. It provides information in terms of patterns associated with lymph nodes and depth extent of the underlying tissues respectively, which is very useful when treating inflammatory diseases such as scleroderma or psoriasis.

Magnetic resonance imaging (MRI)~\cite{tran2007vivo} has also been widely utilized in the examination of pigmented skin lesions. The application of MRI to dermatology has become practice with the use of specialized surface coils that allow higher resolution imaging than standard MRI coils. The application of MRI in dermatology can provide a detailed picture of a tumor and its depth of invasion in relation to adjacent anatomic structures as well as delineate pathways of tumor invasion~\cite{jalil2008multispectral}. For instance, MRI has been used to differentially evaluate malignant melanoma tumors and subcutaneous and pigmented skin of nodular and superficial spreading melanoma~\cite{rajeswari2003evaluation}.

With the development of machine learning, there have been many works using images obtained by digit cameras or smart phones for skin disease diagnosis~\cite{chao2017smartphone,jahan2016comparative}. Though the quality of these images are not as high as these obtained with professional equipments, such as dermatoscopies, excellent diagnosis performance can also be achieved with advanced image processing and analysis methods.

Apart from the above methods, there are a few other imaging acquisition approaches, including Mole Max, Mole Analyzer, real time Raman spectroscopy, electrical impedance spectroscopy, fiber diffraction, and thermal imaging. Due to the limited space, we omit the detailed introduction of these methods here and the readers may refer to related literature if interested.

\subsection{Datasets}
High-quality data has always been the primary requirement of learning reliable algorithms. Particularly, training a deep neural network requires large amount of labeled data. Therefore, high-quality skin disease data with reliable diagnosis labels is significant for the development of advanced algorithms. Three major types of modalities are utilized for skin disease diagnosis, i.e., clinical images, dermoscopy images and pathological images. Specifically, clinical images of skin lesions are usually captured with mobile cameras for remote examination and taken as medical records for patients~\cite{goyal2019artificial}. Dermoscopy images are obtained with high-resolution digital single-lens reflex (DSLR) or smart phone camera attachments. Pathological images, captured by scanning tissue slides with microscopes and digitalized as images, are served as a gold standard for skin disease diagnosis. Recently, many public datasets for skin disease diagnosis tasks have started to emerge. There exists growing trend in the research community to list these datasets for reference. In the following, we present several publicly available datasets for skin disease.

The publicly available PH2 dataset$\footnote{$\textrm{http://www.fc.up.pt/addi/}$}$ of dermoscopy images was built by Mendonca et. al. in $2003$, including $80$ common nevi, $80$ atypical nevi, and $40$ melanomas~\cite{mendoncya2013dermoscopic}. The dermoscopy images were obtained at the Dermatology Service of Hospital Pedro Hispano (Matosinhos, Portugal) under the same conditions through Tuebinger Mole Analyzer system using a magnification of 20x. They are 8-bit RGB color images with a resolution of $768\times560$ pixels. The dataset includes medical annotation of all the images, namely medical segmentation of lesions, clinical and histological diagnoses and the assessment of several dermoscopic criteria (i.e., colors, pigment network, dots/globules, streaks, regression areas, blue-whitish veil). Since the dataset includes comprehensive metadata, it is often utilized as a benchmark dataset for evaluating algorithms for melanoma diagnosis.

Liao~\cite{liao2016deep} built a skin disease dataset for universal skin disease classification from two different resources: Dermnet and OLE. Dermnet is one of the largest publicly available photo dermatology sources~\cite{dermnet}. It contains more than $23,000$ images of skin diseases with various skin conditions and the images are organized in a two-level taxonomy. Specifically, the bottom-level includes images of more than $600$ kind of skin diseases in a fine-grained granularity and the top-level includes images of $23$ kind of skin diseases. Each class of the top-level includes a subcollection of the bottom-level. OLE dataset includes more than $1,300$ images of skin diseases from the New York State Department of Health. The images can be categorized into $19$ classes and each class can be mapped to one of the bottom-level class of the Dermnet dataset. In light of this, Liao~\cite{liao2016deep} labeled the $19$ classes of images from OLE with their top-level counterparts from Dermnet. It should be noted that the images from the above two datasets contain watermarks. To utilize the two datasets, Liao performed two different experiments. One was to train and test CNN models on the Dermnet dataset only, while the other was to train CNN models on the Dermnet dataset and test them on the OLE dataset.

The International Skin Imaging Collaboration (ISIC) aggregated a large-scale publicly available dataset of dermoscopy images~\cite{codella2018skin}. The dataset contains more than $20,000$ images from leading clinical centers internationally, acquired from various devices used at each center. The ISIC dataset was first released for the public benchmark challenge on dermoscopy image analysis in $2016$~\cite{gutman2016skin,marchetti2018results}. The goal of the challenge was to provide a dataset to promote the development of automated melanoma diagnosis algorithms in terms of segmentation, dermoscopic features detection and classification. In $2017$, the ISIC hosted the second term of the challenge with an extended dataset. The extended dataset provides $2,000$ images for training, with masks for segmentation, superpixel masks for dermoscopic feature extraction and annotations for classification~\cite{li2018skin}. The images are categorized into three classes, i.e., melanoma, seborrheic keratosis and nevus. Melanoma is malignant skin tumor while the other two are the benign skin tumors derived from diverse cells. Additionally, the ISIC provides a validation set with extra $150$ images for evaluation.

The HAM10000 (Human Against Machine with $10,000$ training images) dataset released by Tschandl et. al. includes dermoscopy images from diverse populations acquired and stored by different modalities~\cite{tschandl2018ham10000}. The dataset is publicly available through the ISIC archive and consists of $10,015$ dermoscopy images, which are utilized as a training set for testing machine learning algorithms. Cases include a representative collection of all important diagnostic categories in the realm of pigmented lesions. The diagnoses of all melanomas were verified through histopathological evaluation of biopsies, while the diagnoses of nevi were made by either histopathological examination (бл$24\%$), expert consensus (бл$54\%$) or another diagnosis method, such as a series of images that showed no temporal changes (бл$22\%$).

The Interactive Atlas of Dermoscopy (IAD)~\cite{argenziano2002dermoscopy} is a multimedia project for medical education based on a CD-ROM dataset and the dataset includes $2,000$ dermoscopy images and $800$ context images, i.e. non-dermoscopy regular photos. Images in the dataset are labeled as either a melanoma or benign lesion based on pathology report.

The MED-NODE dataset$\footnote{$\textrm{http://www.cs.rug.nl/\~imaging/databases/melanoma\_naevi/}$}$ consists of $70$ melanoma and $100$ naevus images from the digital image archive of the Department of Dermatology of the University Medical Center Groningen (UMCG). It is used for the development and testing of the MED-NODE system for skin cancer detection from macroscopy images~\cite{argenziano2002dermoscopy}.

Dermnet is the largest independent photo dermatology source dedicated to online medical education through articles, photos and videos~\cite{dermnet}. Dermnet provides information on a wide variety of skin conditions through innovative media. It contains over $23,000$ images of skin diseases. Images can be enlarged via a click and located by browsing image categories or using a search engine. The images and videos are available without charge, and users can purchase and license high-resolution copies of images for publishing purposes.

The Dermofit Image Library is a collection of $1,300$ focal high-quality skin lesion images collected under standardized conditions with internal color standards~\cite{dermofit}. The lesions span across ten different classes, including actinic keratosis, basal cell carcinoma, melanocytic nevus, seborrhoeic keratosis, squamous cell carcinoma, intraepithelial carcinoma, pyogenic granuloma, haemangioma, dermatofibroma, and malignant melanoma. Each image has a gold standard diagnosis based on expert opinions (including dermatologists and dermatopathologists). Images consist of a snapshot of the lesion surrounded by some normal skin. A binary segmentation mask that denotes the lesion area is included with each lesion.

The Hallym dataset consists of $152$ basal cell carcinoma images obtained from $106$ patients treated between $2010$ and $2016$ at Dongtan Sacred Heart Hospital, Hallym University, and Sanggye Paik Hospital, Inje University~\cite{han2018classification}.

AtlasDerm contains $10,129$ images of all kinds of dermatology diseases. Samuel Freire da Silva, M.D. created it in homage to The Master And Professor Delso Bringel Calheiros~\cite{atlasDerm}. %For more details about the data, the reader may visit the website: $\textrm{www.atlasdermatologico.com.br}$.

Danderm contains more than $3,000$ clinical images of common skin diseases. This atlas of clinical dermatology is based on photographs taken by Niels K. Veien in a private practice of dermatology~\cite{danderm}. %For more information, please refer to the website: $\textrm{http://www.danderm.dk/}$.

Derm$101$ is an online and mobile resource\footnote{$\textrm{www.derm101.com}$} for physicians and healthcare professionals to learn the diagnosis and treatment of dermatologic diseases~\cite{boer2007derm101}. The resource includes online textbooks, interactive quizzes, peer-reviewed open access dermatology journals, a dermatologic surgery video library, case studies, thousands of clinical photographs and photomicrographs of skin diseases, and mobile applications. %For more information, please refer to the website: $\textrm{www.derm101.com}$.

$7$-point criteria evaluation dataset\footnote{$\textrm{http://derm.cs.sfu.ca}$} includes over $2,000$ dermoscopy and clinical images of skin lesions, with $7$-point checklist criteria and disease diagnosis annotated~\cite{kawahara2018seven}. Additionally, derm$7$pt\footnote{$\textrm{https://github.com/jeremykawahara/derm7pt}$}, a Python module, serves as a starting point to use the dataset. It preprocesses the dataset and converts the data into a more accessible format.

The SD-198 dataset\footnote{$\textrm{https://drive.google.com/file/d/1YgnKz3hnzD3umEYHAgd29n2AwedV1Jmg/view}$} is a publicly available clinical skin disease image dataset. It was built by Sun et al. and includes $6,584$ images from $198$ classes, varying in terms of scale, color, shape and structure~\cite{sun2016benchmark}.

DermIS.net is the largest dermatology information service available on the internet. It offers elaborate image atlases (DOIA and PeDOIA) complete with diagnoses and differential diagnoses, case reports and additional information on almost all skin diseases~\cite{DermIS}.

MoleMap\footnote{$\textrm{http://molemap.co.nz}$} is a dataset that contains $102,451$ images with $25$ skin conditions, including $22$ benign categories and $3$ cancerous categories. In particular, the cancerous categories include melanoma (pink melanoma, normal melanoma and lentigo melanoma), basal cell carcinoma and squamous cell carcinoma~\cite{yi2018unsupervised}. Each lesion has two images: a close-up image taken at a distance of $10$ cm from the lesion (called the macro) and a dermoscopy image of the lesion (called the micro). Images were selected according to four criterion: 1) each image has a disease specific diagnosis (e.g., blue nevus); 2) there are at least $100$ images with the same diagnosis; 3) the image quality is acceptable (e.g., with good contrast); 4) the lesion occupies most of the image without much surrounding tissues.

Asan dataset~\cite{han2018classification} was collected from the Department of Dermatology at Asan Medical Center. It contains $17,125$ clinical images of $12$ types of skin diseases found in Asian people. In particular, the Asan Test dataset containing $1,276$ images is available to be downloaded for research.

The Cancer Genome Atlas~\cite{CancerGenomeAtlas} is one of the largest collections of pathological skin lesion slides that contains $2,860$ cases. The atlas is publicly available to be downloaded for research.

The above publicly available datasets for skin diseases are listed in Table~\ref{dataset}. This may not an exhaustive list for skin disease diagnosis and readers could research the internet for that purpose if interested. From the description of the above skin datasets we can observe that these datasets are usually small in terms of the samples and patients. Compared to the datasets for general computer vision tasks, where datasets typically contain a few hundred thousand and even millions of labeled data, the data sizes for skin disease diagnosis tasks are too small.

\begin{table}
\caption{List of public datasets for skin disease.}\label{dataset}
\center
\scriptsize
%\resizebox{\textwidth}{8cm}{
\begin{tabular}{p{3.8cm}cp{4.5cm}}
\hline
{\bf Dataset}                                     &  {\bf No. of images}                & {\bf Type of skin disease}        \\
\hline
PH$2$ dataset~\cite{mendoncya2013dermoscopic}     & 200                                 & Common nevi, melanomas, atypical nevi\\
~\cite{liao2016deep}                              & $>$ 3,600                           & 19 classes  \\
ISIC~\cite{codella2018skin}                       & $>$ 20,000                               & Melanoma, seborrheic keratosis, benign nevi\\
HAM10000~\cite{tschandl2018ham10000}              & 10,015                              & Important diagnostic categories of pigmented lesions\\
IAD~\cite{argenziano2002dermoscopy}               & 2,000                               & Melanoma and benign lesion\\
MED-NODE dataset~\cite{argenziano2002dermoscopy}  & 170                                 & Melanoma and nevi\\
Dermnet~\cite{dermnet}                            & 23,000                              & All kinds of skin diseases \\
Dermofit Image Library~\cite{dermofit}            & 1,300                               & $10$ different classes \\
Hallym dataset~\cite{han2018classification}       & 152                                 & Basal cell carcinoma \\
AtlasDerm~\cite{atlasDerm}                        & 10,129                              & All kinds of skin diseases \\
Danderm~\cite{danderm}                            & 3,000                               & Common skin diseases \\
Derm101~\cite{boer2007derm101}                    & Thousands                           & All kinds of skin diseases \\
7-point criteria evaluation dataset~\cite{kawahara2018seven} & $>$ 2,000                & Melanoma and non-melanoma \\
SD-198 dataset~\cite{sun2016benchmark}            & 6,584                               & $198$ classes \\
DermIS~\cite{DermIS}                              & Thousands                           & All kinds of skin diseases\\
MoleMap~\cite{yi2018unsupervised}                 & $102,451$                           & $22$ benign categories and $3$ cancerous categories\\
Asan dataset~\cite{han2018classification}         & $17,125$                            & $12$ types of skin diseases found in Asian people\\
The Cancer Genome Atlas~\cite{CancerGenomeAtlas}  & $2,860$                             & Common skin diseases \\
\hline
\end{tabular}%}
\end{table}

\section{Deep learning}

In the area of machine learning, people design models to enable computers to solve problems by learning from experiences. The aim is to develop models that can be trained to produce valuable results when fed with new data. Machine learning models transform their input into output with statistical or data-driven rules derived from large numbers of examples~\cite{esteva2019guide}. They are tuned with training data to obtain accurate predictions. The ability of generalizing the learned expertise to make correct predictions for new data is the main goal of the models. The generalization ability of a model is estimated during the training process with a separate validation dataset and utilized as feedback for further tuning. Then the fully tuned model is evaluated on a testing dataset to investigate how well the model makes predictions for new data.

There are several types of machine learning models, which can be classified into three categories, i.e., supervised learning, semi-supervised learning and unsupervised learning models, according to how the data is used for training a model. In supervised learning, a model is trained with labeled or annotated data and then used to make predictions for new, unseen data. It is called supervised learning since the process of learning from the training data can be considered as a teacher supervising the learning process. Most of machine learning models adopt supervised learning. For instance, classifying skin lesions into classes of ``benign" or ``malignant" is a task using supervised learning~\cite{mahbod2019fusing}. By contrast, in unsupervised learning, the model is aimed to discover the underlying distribution or structure in the data in order to learn more about the data without guidance. Clustering~\cite{hartigan1979algorithm} is a typical unsupervised learning model. Problems where you have large amounts of data and only some of the data is labeled are called semi-supervised learning problems~\cite{zhu2005semi}. These problems sit in between both supervised learning and unsupervised learning. Actually, many real-world machine learning problems, especially medical image processing, fall into this type. It is because that labeling large amounts of data can be expensive or time-consuming. By contrast, unlabeled data is more common and easy to obtain.

Machine learning has a long history and can be split into many subareas. Particularly, deep learning is a branch of machine learning and has been popular in the past few years. Previously, designing a machine learning algorithm required domain information or human engineering to extract meaningful features that can be a representation of data and input to an algorithm for pattern recognition. However, a deep learning model consisting of multiple layers is a kind of representation learning method that transforms the input raw data into needed representation for pattern recognition without much human interference. The layers in a deep learning architecture are arranged sequentially and composed of large numbers of predefined, nonlinear operations, such that the output of one layer is input to the next layer to form more complex and abstract representations. In this way, a deep learning architecture is able to learn complex functions. With the ability of running on specialized computational hardware, deep learning models adapt large-scale data and can be optimized with more data continually. As a result, deep learning algorithms outperform most of conventional machine learning algorithms in many problems. People have witnessed the huge development of deep learning algorithms and their extensive applications in various tasks, such as object classification~\cite{krizhevsky2012imagenet,eitel2015multimodal,qi2017pointnet}, machine translation~\cite{wu2016google,zhou2016deep} and speech recognition~\cite{chorowski2015attention,amodei2016deep,afouras2018deep}. Particularly, healthcare and medicine benefit a lot from the prevalence of deep learning due to the huge volume of medical data~\cite{esteva2019guide,he2019practical}. Three major factors have contributed the success of deep learning for solving complex problems of modern society, including: 1) availability of massive training data. With the ubiquitous digitization of information in recent world, public sufficiently large volumes of data is available to train complex deep learning models; 2) availability of powerful computational resources. Training complex deep learning models with massive data requires immense computational power. Only the availability of powerful computational resources, especially the improvements in graphic processing unit (GPU) performance and the development of methods to use the GPU for computation, in recent times fulfills such requirements; 3) availability of deep learning frameworks. People in diverse research communities are more and more willing to share their source codes on public platforms. Easy access to deep learning algorithm implementations, such as GoogLeNet~\cite{szegedy2015going}, ResNet~\cite{he2016deep}, DenseNet~\cite{huang2017densely} and SENet~\cite{hu2018squeeze}, has accelerated the speed of applying deep learning to practical tasks.

Commonly, deep learning models are trained in a supervised way, i.e., the datasets for training contain data points (e.g., images of skin diseases) and corresponding labels (e.g., ``benign" or ``malignant") simultaneously. However, data labels are limited for healthcare data since labeling large numbers of data is expensive and difficult. Recently, semi-supervised and unsupervised learning have attracted much attention to alleviate the issues caused by limited labeled data. There have been many excellent reviews and surveys of deep learning~\cite{lecun2015deep,goodfellow2016deep,bengio2017deep,litjens2017survey} and interested readers can refer them for more details.

In the following, we briefly introduce the essential part of deep learning, aiming to provide a useful guidance to the area of skin disease diagnosis that are currently influenced by deep learning.

\subsection{Neural networks}

Neural networks are a type of learning algorithm that formulates the basis of most deep learning algorithms. A neural network consists of neurons or units with activation $z$ and parameters $\Theta=\{\omega,\ \beta\}$, where $\omega$ is a set of weights and $\beta$ a set of biases. The activation $z$ is expressed as a linear combination of the input $\bm{x}$ to the neuron and parameters, followed with an element-wise nonlinear activation function $\sigma(\cdot)$:
\begin{equation}
	z=\sigma(\bm{w}^{T}\bm{x}+b),
\end{equation}
where $\bm{w}\in\omega$ is the weight and $b\in\beta$ is the bias. Typical activation functions for neural networks include the sigmoid function and hyperbolic tangent function. Particularly, the multi-layer perceptrons (MLPs) are the most well-known neural networks, containing multiple layers of this kind of transformations:
\begin{equation}
	f(\bm{x}; \Theta)=\sigma(\bm{W}^{L}(\sigma(\bm{W}^{L-1}\cdots\sigma(\bm{W}^{0}+b^{0})\cdots+b^{L-1})+b^{L}),
\end{equation}
where $\bm{W}^{n},\ n=1,2,\cdots,L$ is a matrix consisting of rows $\bm{w}^{k},\ k=1,2,\cdots,n_{c}$ which are associated with the $k$-th activation in the output, $L$ indicates the total number of layers and $n_{c}$ indicates the number of nodes at the $n$-th layer. The layers between the input and output layers are often called as ``hidden" layers. When a neural network contains multiple layers, then we say it is a deep neural network. Hence, we have the term ``deep learning".

\begin{figure}
\center
\includegraphics[width=0.6\textwidth]{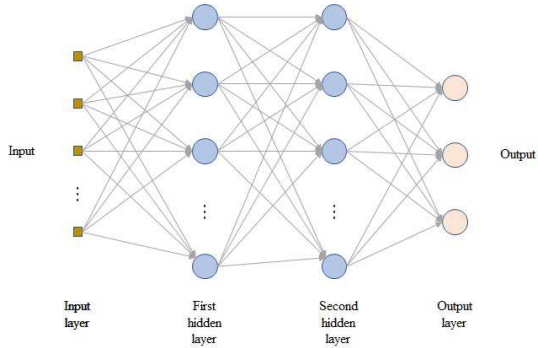}
\caption{An example of a $4$-layer MLPs.} \label{mlps}
\end{figure}

Commonly, the activations of the final layer of a network are mapped to a distribution over classes $p(y|\bm{x};\Theta)$ via a softmax function~\cite{litjens2017survey}:
\begin{equation}
p(y|\bm{x};\Theta)=softmax(\bm{x};\Theta)=\frac{e^{(\bm{w^{L}_{c}})^{T}\bm{x}+b^{L}_{c}}}{\sum^{C}_{c=1}e^{(\bm{w^{L}_{c}})^{T}\bm{x}+b^{L}_{c}}},
\end{equation}
where $\bm{w}^{L}_{c}$ indicates the weight that produces the output node corresponding to class $c$. An example of a $4$-layer MLPs is illustrated in Fig.~\ref{mlps}.

Currently, stochastic gradient descent (SGD) is the most popular method used for tuning the parameters $\Theta$ for a specific dataset. In SGD, a mini-batch, i.e., a small subset of the dataset, is utilized for the gradient update instead of the whole dataset. Tuning the parameters is to minimize the negative log-likelihood:
\begin{equation}
\arg min_{\Theta}-\sum^{N}_{n=1}log(p(y_{n}|\bm{x_{n}};\Theta)).
\end{equation}
Practically, one can design the loss function according to the specific tasks. For example, the binary cross-entropy loss is used for two-class classification problems and the categorical cross-entropy loss for multi-class classification problems.

For a long time, people considered that deep neural networks (DNNs) were hard to train. Major breakthrough was made in $2006$ when researchers showed that training DNNs layer-by-layer in an unsupervised way (pretraining), followed with a supervised fine-tuning of the stacked layers, could obtain promising performance~\cite{hinton2006reducing,bengio2007greedy,hinton2006reducing}. Particularly, the two popular networks trained in such a manner are stacked autoencoders (SAEs)~\cite{vincent2008extracting} and deep belief networks (DBNs)~\cite{hinton2009deep}. However, such techniques are complicated and require many engineering tricks to obtain satisfying performance.

Currently, most popular architectures are trained end-to-end in a supervised way, which greatly simplifies the training processes. The most prevalent models are convolutional neural networks (CNNs)~\cite{krizhevsky2012imagenet} and recurrent neural networks (RNNs)~\cite{mikolov2010recurrent}. In particular, CNNs are extensively applied in the field of medical image analysis~\cite{shin2016deep,milletari2016v,lakhani2017deep}. They are powerful tools for extracting features from images and other structured data. Before it became possible to utilize CNNs efficiently, features were typically obtained by handcrafted engineering methods or less powerful traditional machine learning models. The features learned from the data directly with CNNs show superior performance compared with the handcrafted features. There are strong preferences about how CNNs are constructed, which can benefit us to understand why they are so powerful. Therefore, we give a brief introduction to the building blocks of CNNs in the following.

\subsection{Convolutional neural networks}

\begin{figure}
\center
\includegraphics[width=0.8\textwidth]{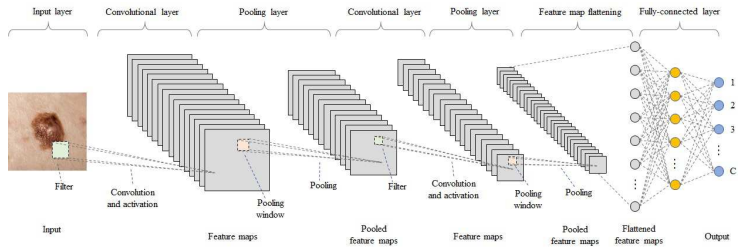}
\caption{An illustration of a typical CNN.} \label{CNN}
\end{figure}

One can utilize the feedforward neural networks discussed above to process images. However, having connections between all the nodes in one layer and all the nodes in the next layer is quite inefficient. A careful pruning of the connections based on the structure of images can lead to better performance with high efficiency. CNNs are special kind of neural networks that preserve the spatial relationships in the data with very few connections between layers. CNNs are able to extract meaningful representations from input data, which are particularly appropriate for image-oriented problems. A CNN consists of multiple layers of convolutions and activations, with pooling layers interspersed between different convolution layers. It is trained via backpropagation and SGD similar with the standard neural networks. Additionally, a CNN typically includes fully-connected layers at the end of the architecture to produce the output. A typical CNN is demonstrated in Fig.~\ref{CNN}.

\subsubsection{Convolutional layers}
\begin{figure}
\center
\includegraphics[width=0.6\textwidth]{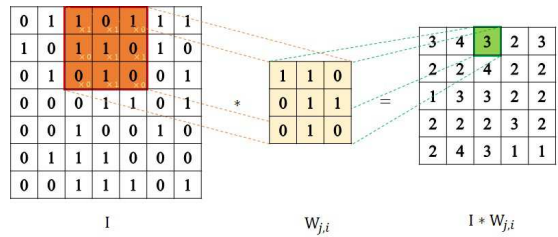}
\caption{An illustration of a 2D convolution operation.} \label{conv}
\end{figure}

In the convolutional layers, the output activations of the previous layer are convolved with a set of filters represented with a tensor $\bm{W}_{j,i}$, where $j$ is the filter number and $i$ is the layer number. Fig.~\ref{conv} demonstrates a 2D convolution operation. The operation involves moving a small window of size $3\times3$ over a 2D grid (e.g., an image or a feature map) in a left-to-right and up-to-down order. At each step, the corresponding elements of the window and grid are multiplied and summed up to obtain a scalar value. With all the obtained values, another 2D grid is produced, referred as feature map in a CNN. By having each filter share the same weights across the whole input domain, much less number of weights is needed. The motivation of the weight-sharing mechanism is that the features appearing in one part of the image are likely to appear in other parts as well~\cite{lundervold2019overview}. For example, if you have a filter that can detect vertical lines, then it can be utilized to detect lines wherever they appear. Applying all the convolutional filters to all locations of the input results in a set of feature maps.

\subsubsection{Activation layers}
The outputs from convolutional layers are fed into a nonlinear activation function, which makes it possible for the neural network to approximate almost any nonlinear functions~\cite{leshno1993multilayer}. It should be noted that a multi-layer neural network constructed with linear activation functions can only approximate linear functions. The most common activation function is rectified linear units (ReLU), which is defined as $\textrm{ReLU}(z)=\max(0,z)$. There have many variants of ReLU, such as leaky ReLU (LeakyReLU)~\cite{xu2015empirical} and parametric ReLU (PReLU)~\cite{he2015delving}. The outputs of the activation functions are new tensors and we call them feature maps.

\subsubsection{Pooling layers}
The feature maps output by the activation layers are then typically pooled in the pooling layers. The pooling operations are performed on a small region (e.g., a square region) of the input feature maps and only one single value is obtained with certain scheme. The common schemes utilized to compute the value are max function (max pooling) and average function (average pooling). A small shift in the input image will lead to small changes in the activation maps; however, the pooling operation enables the CNNs to have the translation invariance property. Another way to obtain the same downsampling effect as the pooling operation is to perform convolution with a stride larger than one pixel. Researches have shown that removing pooling layer could simplify the networks without sacrificing performances~\cite{springenberg2014striving}.

Besides the above building blocks, other important elements in many CNNs include dropout and batch normalization. Dropout~\cite{srivastava2014dropout} is a simple but powerful tool to boost the performance of CNNs. Averaging the performance of several models in an ensemble one tends to obtain better performance than any of the single model. Dropout performs similar averaging operation based on the stochastic sampling of neural networks. With dropout, one randomly removes neurons in the networks during training process, ending up utilizing slightly different networks for each batch of the training data. As a result, the weights of the networks are tuned based on optimizing multiple different variants of the original networks. Batch normalization is often placed after the activation layers and produces normalized feature maps by subtracting the mean and dividing with the standard deviation for each training batch~\cite{ioffe2015batch}. With batch normalization, the networks are forced to keep their activations being zero mean and unit standard deviation, which works as a network regularization. In this way, the networks training process can be speeded up and less dependent on the careful parameter initialization.

When designing new and more advanced CNN architectures, these components are combined together in a more complicated way and other ingredients can be added as well. To construct a specific CNN architecture for a practical task, there are a few factors to be considered, including understanding the tasks to be solved and the requirements to be satisfied, finding out how to preprocess the data before input to a network, and making full use of the available budget of computation. In the early days of modern deep learning, people designed networks simply with the combination of the above building blocks, such as LeNet~\cite{lecun1990handwritten} and AlexNet~\cite{krizhevsky2012imagenet}. Later, the architectures of networks became more and more complex in a way that they were built based on the ideas and insights of previous models. Table~\ref{Architetures} and~\ref{Architetures-2} demonstrate a few popular deep network architectures, hoping to show how the building blocks can be combined to create networks with excellent performances. These DNNs are typically implemented in one or more of a small number of deep learning frameworks that are introduced in detail in the next section. Thanks to the software development platform, such as GitHub, the implementation of large numbers of DNNs with the main deep learning frameworks have been made publicly accessible, which makes it easier for people to reproduce or reuse these models.

\begin{table}
\caption{A few popular deep network architectures (part $1$).}\label{Architetures}
\centering
\scriptsize
\begin{tabular}{p{1.4cm}<{\centering}ccp{8.5cm}}
\hline
{\bf Architecture}  & {\bf Year}      &  {\bf Reference}                & {\bf Description}         \\
\hline
LeNet               & 1990     & \cite{lecun1990handwritten}     & Proposed by Yann LeCun to solve the task of handwritten digit recognition. Since then, the basic architecture of CNN has been fixed: convolutional layer, pooling layer and fully-connected layer.                          \\
AlexNet             & 2012     & \cite{krizhevsky2012imagenet}   & Considered as one of the most influential works in the field of computer vision since it has spurred many more papers utilizing CNN and GPUs to accelerate deep learning~\cite{deshpande20189}. The building blocks of the network include convolutional layers, ReLU activation function, max-pooling and dropout regularization. In addition, the authors split the computations on multiple GPUs to make training faster. It won the $2012$ ILSVRC competition by a huge margin.\\
VGG-nets            & 2014     & \cite{simonyan2014very}         & Proposed by the Visual Geometry Group (VGG) of the Oxford University and won the first place for the localization task and the second place for the classification task in the 2014 ImageNet competition. VGG-nets can be seen as a deeper version of AlexNet. They adopt a pretraining method for network initialization: train a small network first and ensure that this part of the network is stable, and then go deeper gradually based on this.    \\
GoogLeNet           & 2015     & \cite{szegedy2015going}         & Defeated VGG-nets in the classification task of 2014 ImageNet competition and won the championship. Different from networks like AlexNet, VGG-nets which rely solely on deepening networks to improve performance, GoogLeNet presents a novel network structure whiling deepens the network ($22$ layers). A inception structure replaces the traditional operations of convolution and activation. This idea was first proposed by the Network in Network~\cite{lin2013network}. In the inception structure, multiple filters of diverse sizes are performed to the input and the corresponding results are concatenated. This multi-scale processing enables the network to extract features at different scales efficiently.     \\
ResNet              & 2016     & \cite{he2016deep}               & Introduces the residual module, which makes it easier to train much deeper networks. The residual module consists of a standard pathway and a skip connection, providing options to the network to simply copy the activations from one residual module to the next module. In this way, information can be preserved when data goes through the layers. Some features are best extracted with shallow networks, while others are best extracted with deeper ones. Residual modules enable the network to include both cases simultaneously, which performs similarly as ensemble and increases the flexibility of the network. The $152$-layer ResNet won the 2015 ILSVRC competition, and the authors also successfully trained a version with $1,001$ layers.  \\
ResNext             & 2017     & \cite{xie2017aggregated}        & Built based on ResNet and GoogLeNet by incorporating inception modules between skip connections. \\
DenseNet            & 2017     & \cite{huang2017densely}         & A neural network with dense connections. In this network, there is a direct connection between any two layers. That is to say, the input of each layer is the union of the outputs of all previous layers, and the feature map learned by the layer is also directly transmitted to all layers afterwards. In this way, the network mitigates the problem of gradient disappearance, enhances feature propagation, encourages feature reuse, and greatly reduces the amount of parameters.     \\
\hline
\end{tabular}
\end{table}

\begin{table}
\caption{A few popular network architectures (part $2$).}\label{Architetures-2}
\centering
\scriptsize
\begin{tabular}{p{1.4cm}<{\centering}ccp{8.5cm}}
\hline
{\bf Architecture}  & {\bf Year}      &  {\bf Reference}                & {\bf Description}         \\
\hline
SENets              & 2018     & \cite{hu2018squeeze}            & Squeeze-and-Excitation (SE) network, which is built by introducing SE modules into existing networks. The SE modules are trained to weight the feature maps channel-wise. Consequently, the SENets are able to model spatial and channel information separately, enhancing the model capacity with negligible increase in computational costs.     \\
NASNet              & 2018     & \cite{zoph2018learning}         & A CNN architecture designed by AutoML which is a reinforcement learning approach used for neural network architecture searching~\cite{bello2017neural}. A controller network proposes architectures aimed to perform at a specific level for a specific task, and learns to propose better models by trial and error. NASNet was built based on CIFAR-$10$ with relatively modest computation requirements, outperforming all previous human-designed networks in the ILSVRC competition.       \\
GAN                & 2014     & \cite{goodfellow2014generative} & Generative adversarial network (GAN) was proposed by Goodfellow et al. in $2014$ and developed rapidly in recent years. A GAN consists of two networks that compete against each other. The generative network $G$ creates samples to make the discriminative network $D$ think they come from the training data rather than the generative network. The two networks are trained alternatively, where $G$ aims to maximize the probability that $D$ makes a mistake while $D$ aims to obtain high classification accuracy. There have been a variety of variants (DCGANs~\cite{radford2015unsupervised}, CycleGAN~\cite{zhu2017unpaired}, SAGAN~\cite{zhang2018self} etc.) so far and they developed into a subarea of machine learning.\\
U-net               & 2015     & \cite{ronneberger2015u}         & A very popular and successful network for $2$-D medical image segmentation. Fed with an image, the network first downsamples the image with a traditional CNN architecture and then upsamples the resulting feature maps through a serial of transposed convolution operations to the same size as the original input image. Additional, there have skip connections between the downsampling and upsampling counterparts. \\
%YOLO                     & \cite{redmon2016you}            &         \\
Faster R-CNN        & 2015       & \cite{ren2015faster}         & The faster region-based convolutional network was built based on the previous Fast R-CNN~\cite{girshick2015fast} for object detection. The major contribution of the method is to develop a region proposal network (RPN) to further reduce the region proposal computation time. The region proposal is nearly cost-free, and therefore the object detection system can run at near real-time frame rates.      \\
Mask R-CNN          & 2017     & \cite{he2017mask}               & Extends Faster R-CNN by adding a branch for predicting an object mask in parallel with the existing branch for bounding box recognition. The method can generate a high-quality segmentation mask for each instance while efficiently detect the objects in the image. Mask R-CNN is simple to train and adds only a small overhead to Faster R-CNN. It outperforms all previous, single-model entries on all three tracks of the COCO suite of challenges.        \\
\hline
\end{tabular}
\end{table}

\section{Deep learning frameworks}

With the prevalence of deep learning, there are several open source deep learning frameworks aiming to simplify the implementation of complex and large-scale deep learning models. Deep learning frameworks provide building blocks for designing, training and validating DNNs with high-level programming interfaces. Thus, people can implement complex models like CNNs conveniently. In the following, we present a brief introduction to popular deep learning frameworks.%, including TensorFlow, Keras, PyTorch, Caffe, Sonnet, and MXNet.

TensorFlow~\cite{abadi2016tensorflow} was developed by researchers and engineers from the Google Brain team. It is by far the most popular software library in the field of deep learning (though others are catching up quickly). One of the biggest reasons accounting for the popularity of TensorFlow is that it supports multiple programming languages, such as Python, C++ and R, to build deep learning models. It is handy for creating and experimenting with deep learning architectures. In addition, its formulation is convenient for data (such as inputting graphs, SQL tables, and images) integration. Moreover, it provides proper documentations and walkthroughs for guidance. The flexible architecture of TensorFlow makes it easy for people to run their deep learning models on one or more CPUs and GPUs. It is backed by Google, which guarantees that it will stay around for a while. Therefore, it makes sense to invest time and resources to use it.

Keras~\cite{chollet2015keras} is written with Python and can run on top of TensorFlow (as well as CNTK and Theano). The interface of TensorFlow can be a little challenging for new users since it is a low-level library, and therefore new users may find it hard to understand certain implementations. By contrast, Keras is a high-level API, developed with the aim of enabling fast experimentation. It is designed to minimize the user actions and make it easy to understand models. However, this strategy makes Keras a less configurable environment than low-level frameworks. Even so, Keras is appropriate for deep learning beginners that are unable to understand complex models properly. If you want to obtain results quickly, Keras will automatically take care of the core tasks and produce outputs. It runs seamlessly on multiple CPUs and GPUs.

PyTorch~\cite{paszke2017automatic}, released by Facebook, is a primary software tool for deep learning after Tensorflow. It is a port to the Torch deep learning framework that can be used for building DNNs and executing tensor computations. Torch is a Lua-based framework while PyTorch runs on Python. PyTorch is a Python package that offers Tensor computations. Tensors are multidimensional arrays like ndarrays in numpy that can run on GPUs as well. PyTorch utilizes dynamic computation graphs. Autograd package of PyTorch builds computation graphs from tensors and automatically computes gradients. Instead of predefined graphs with specific functionalities, PyTorch offers us a framework to build computation graphs as we go, and even change them during runtime. This is valuable for situations where we do not know how much memory is needed for creating a DNN. The process of training a neural network is simple and clear, and PyTorch contains many pretrained models.

Caffe~\cite{jia2014caffe} is another popular open source deep learning framework designed for image processing. It was developed by Yangqing Jia during his Ph.D. at the University of California, Berkeley. First of all, it should be noted that its support for recurrent networks and language modeling is not as great as the above three frameworks. However, Caffe presents advantages in terms of the speed of processing and learning from images. Caffe provides solid support for multiple interfaces, including C, C++, Python, MATLAB as well as traditional command line. Moreover, the Caffe Model Zoo framework allows us to utilize pretrained networks, models and weights that can be used to solve deep learning tasks.

Sonnet~\cite{sonnet} is a deep learning framework built based on top of TensorFlow. It is designed to construct neural networks with complex architectures by the world-famous company DeepMind. The idea of Sonnet is to construct primary Python objects corresponding to a specific part of the neural network. Furthermore, these objects are independently connected to computational TensorFlow graphs. Separating the process of creating objects and associating them with a graph simplify the design of high-level architectures. The main advantage of Sonnet is that you can utilize it to reproduce the research demonstrated in the papers of DeepMind. In summary, it is a flexible functional abstraction tool that is absolutely a worthy opponent for TensorFlow and PyTorch.

MXNet is a highly scalable deep learning framework that can be applied on a wide variety of devices~\cite{chen2015mxnet}. Although it is not as popular as TensorFlow, the growth of MXNet is likely to be boosted by becoming an Apache project. The framework initially supports a large number of programming languages, such as C++, Python, R, Julia, JavaScript, Scala, Go and even Perl. The framework is very efficient for parallel computing on multiple GPUs and machines. MXNet has detailed documentation and is easy to use with the ability to choose between imperative and symbolic programming styles, making it a great candidate for both beginners and experienced engineers.

Besides the above six frameworks, there have other less popular but useful deep learning frameworks, such as Microsoft Cognitive Toolkit, Gluon, Swift, Chainer, DeepLearning4J, Theano, PaddlePaddle and ONNX. Due to the limitation of space, we cannot detail them all here. If interested, readers may find more related information by searching the internet. Note that all the frameworks are built on top of NVIDIA's CUDA platform and the cuDNN library, and are open source and under active development.

\section{Evaluation metrics}

\subsection{Segmentation tasks}
For segmentation tasks, the most common evaluation metric is Intersection-over-Union (IoU), also known as Jaccard Index. IoU is to measure the overlap between the segmented area predicted by algorithms and that of the ground-truth, i.e.,
\begin{equation}
IoU=\frac{Area\ of\ overlap}{Area\ of\ union}
\end{equation}
wherer $Area\ of\ overlap$ indicates the overlap of the segmented area predicted by algorithms and that of the ground-truth, and $Area\ of\ union$ indicates the union of the two items. The value of IoU ranges from $0$ to $1$ and higher value means better performance of the algorithms.

Besides IoU, the following indices are utilized for evaluating a segmentation algorithm as well.

Pixel-level accuracy:
\begin{equation}
AC=\frac{TP+TN}{TP+FP+TN+FN}
\end{equation}
where TP, TN, FP, FN denote true positive, true negative, false positive and false negative at the pixel level, respectively. Pixel values above $128$ are considered positive, and pixel values below $128$ are considered negative.

Pixel-level sensitivity:
\begin{equation}
SE=\frac{TP}{TP+FN}
\end{equation}

Pixel-level specificity:
\begin{equation}
SP=\frac{TN}{TN+FP}
\end{equation}

Dice Coefficient:
\begin{equation}
DI=\frac{2TP}{2TP+FN+FP}
\end{equation}

%Jaccard Index:
%\begin{equation}
%JA=\frac{TP}{TP+FN+FP}
%\end{equation}

\subsection{Classification tasks}

For classification tasks, common evaluation metrics include accuracy, sensitivity and specificity, which are the same with those defined for segmentation tasks. However, metrics are measured at the whole image level instead of the pixel level. In addition, the area under the receiver operation characteristic (ROC) curve (AUC) and precision are also common measurements.

The AUC measures how well a parameter can be distinguished between two diverse groups and is computed by taking the integral of true positive rate regarding the false positive rate:
\begin{equation}
AUC=\int^{1}_{0}t_{pr}(f_{pr})\delta f_{pr}
\end{equation}

Precision is defined as the following:
\begin{equation}
PREC=\frac{TP}{TP+FP}
\end{equation}

\section{Skin disease diagnosis with deep learning}

\begin{figure}
\center
\includegraphics[width=0.6\textwidth]{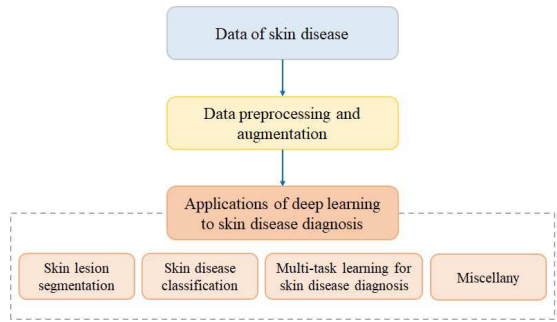}
\caption{The taxonomy of literature review of skin disease diagnosis with deep learning.} \label{app}
\end{figure}

Given the popularity of deep learning, there have been numerous applications of deep learning methods in the tasks of skin disease diagnosis. In this section, we review the existing works in skin disease diagnosis that exploit the deep learning technology. From a machine learning perspective, we first introduce the common data preprocessing and augmentation methods utilized in deep learning and then present the review of existing literature on applications of deep learning in skin disease diagnosis according to the type of tasks. The taxonomy of the literature review of this section is illustrated in Fig.~\ref{app}.

\subsection{Data preprocessing and augmentation}

\subsubsection{Data preprocessing}

Data preprocessing plays an important role in skin disease diagnosis with deep learning. Since there is a huge variation in image resolutions of skin disease datasets (e.g., ISIC, PH2 and AtlasDerm) and deep networks commonly receive inputs with certain square sizes (e.g., $224\times224$ and $512\times512$), it is necessary to crop or resize the images from these datasets to adapt them to deep learning networks. It should be noted that resizing and cropping images directly into required sizes might introduce object distortion or substantial information loss~\cite{zheng2016good,yu2018melanoma}. Feasible methods to resolve this issue is to resize images along the shortest side to a uniform scale while maintaining the aspect ratio. Typically, images are normalized by subtracting the mean value and then divided by the standard deviation, which are calculated over the whole training subset, before fed into a deep learning network. There have works~\cite{rastgoo2015automatic,yu2018melanoma} reported that subtracting a uniform mean value does not well normalize the illumination of individual images since the lighting, skin tones and viewpoints of skin disease images may vary greatly across a dataset. To address this issue, Yu et al.~\cite{yu2018melanoma} normalized each image by subtracting it with channel-wise mean intensity values calculated over the individual image. The experimental results in their paper showed that simply subtracting a uniform mean pixel value will decrease the performance of a deep network. In addition, for more accurate segmentation and classification, hair or other unrelated stuffs should be removed from skin images with algorithms including thresholding methods~\cite{vala2013review,huang2008new}, morphological methods~\cite{parvati2008image}, and deep learning algorithms~\cite{ronneberger2015u,badrinarayanan2017segnet,chen2017deeplab}.

\subsubsection{Data augmentation}
As is known that large numbers of data are usually required for training a deep learning network to avoid overfitting and achieve excellent performances. Unfortunately, many applications, such as skin disease diagnosis, can hardly have access to massive labeled training data. In fact, limited data are common in the field of medical image analysis due to the rarity of disease, patient privacy, the requirement of labeling by medical experts and the high cost to obtain medical data~\cite{shorten2019survey}. To alleviate this issue, data augmentation, indicating artificially transforming original data with some appropriate methods to increase the amount of available training data, are developed. With feasible data augmentation, one can enhance the size and quality of the available training data. With additional data, deep learning architectures are able to learn more significant properties, such as rotation and translation invariance.

Popular data augmentation methods include geometric transformations (e.g., flip, crop, translation, and rotation), color space augmentations, kernel filters, mixing images, random erasing, feature space augmentation, adversarial training, generative adversarial networks, neural style transfer, and meta-learning~\cite{shorten2019survey}. For example, Al-Masni et al.~\cite{al2018skin} augmented training data by rotating all of the $4,000$ dermoscopy images with angles of $0^{\circ}$, $90^{\circ}$, $180^{\circ}$ and $270^{\circ}$. In this way, overfitting was reduced and robustness of deep networks was improved. Yu et al.~\cite{yu2018melanoma} rotated each image by angles of $0^{\circ}$, $90^{\circ}$ and $180^{\circ}$, and then performed random pixel translation (with a shift between $-10$ and $10$ pixels) to the rotated images. Significant improvement was achieved with data augmentation in their experiments on the ISIC skin dataset. Detailed discussion on data augmentation is beyond the scope of this paper and readers may refer to the work by Shorten et al.~\cite{shorten2019survey} for more information.

\subsection{Applications of deep learning in skin disease diagnosis}
\subsubsection{Skin lesion segmentation}

\begin{figure}
\center
\includegraphics[width=0.8\textwidth]{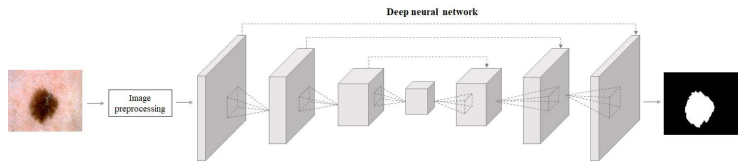}
\caption{The workflow of a typical skin disease segmentation task.} \label{workflow_seg}
\end{figure}

Segmentation aims to divide an image into distinct regions that contain pixels with similar attributes. Segmentation is significant for skin disease diagnosis since it avails clinicians to perceive the boundaries of lesions. The success of image analysis depends on the reliability of segmentation, whereas a precise segmentation of an image is generally challenging. Manual boarder detection considers the quandary caused by collision of tumors, wherein there is proximity of lesions of more than one types. Therefore, higher caliber knowledge of lesion features should be taken into account~\cite{celebi2009lesion}. Particularly, the morphological differences in appearance of skin lesions bring more difficulties to skin diseases segmentation. The foremost reason is that a relatively poor contrast between the mundane and skin lesion exists. Other reasons that make the segmentation difficult include variations in skin tones, presence of artifacts such as hair, ink, air bubbles, ruler marks, non-uniform lighting, physical location of lesions and lesion variations in respect to color, texture, shape, size and location in the image~\cite{celebi2015state,pathan2018techniques}. These factors should be considered when designing a segmentation algorithm for skin disease images. Generally, effective image preprocessing should be adopted to eliminate the impact of these factors before images are input to segmentation algorithms~\cite{liao2016deep,chang2017skin}. In the past few years, deep learning has been extensively applied to image segmentations for skin diseases and achieved promising performance~\cite{yu2016automated,yuan2017automatic,unver2019skin,badrinarayanan2017segnet,peng2019segmentation}. The workflow of a typical skin disease segmentation task is illustrated in Fig.~\ref{workflow_seg}.

Fully convolutional neural network with an encoder-decoder architecture (e.g., fully convolutional network (FCN)~\cite{long2015fully} and SegNet~\cite{badrinarayanan2017segnet}) was one of the earliest deep learning models proposed for semantic image segmentation. Particularly, deep learning models based on FCN have been used for skin lesion segmentation. For instance, Attia et al.~\cite{attia2017skin} proposed a network combining a FCN with a long short term memory (LSTM)~\cite{hochreiter1997long} to perform segmentation for melanoma images. The method did not require any preprocessing to the input images and achieved state-of-the-art performances with an average segmentation accuracy of $0.98$ and Jaccard index of $0.93$ on the ISIC dataset. The authors found that the hybrid method utilizing RNN and CNN simultaneously was able to outperform methods that rely on CNN only. Bi et al.~\cite{bi2017dermoscopic} proposed a FCN based method to automatically segment skin lesions from dermoscopy images. Specifically, multiple embedded FCN stages were proposed to learn important visual characteristics of skin lesions and these features were combined together to segment the skin lesion accurately. Goyal et al.~\cite{goyal2017multi} proposed a multi-class segmentation method based on FCN for benign nevi, melanoma and seborrhoeic keratoses images. The authors tested the method on the ISIC dataset and obtained dice coefficient indices of $55.7\%$, $65.3\%$, and $78.5\%$ for the $3$ classes respectively. Phillips et al.~\cite{phillips2019segmentation} proposed a novel multi-stride FCN architecture for segmentation of prognostic tissue structures in cutaneous melanoma using whole slide images. The weights of the proposed multi-stride network were initiated with multiple networks pretrained on the PascalVOC segmentation dataset and fine-tuned on the whole slide images. Results showed that the proposed approach had the possibility to achieve a level of accuracy required to manually perform the Breslow thickness measurement.

The well-known neural network, U-net~\cite{ronneberger2015u}, was proposed for medical image segmentation in $2015$. The network was constructed based on FCN, and its architecture has been modified and extended to many works that yielded better segmentation results~\cite{cciccek20163d,oktay2018attention}. Naturally, there have been several works applying U-net to the task of skin lesion segmentation. Chang et al.~\cite{chang2017skin} implemented U-net to segment dermoscopy images of melanoma. Then both the segmented images and original dermoscopy images were input to a deep network consisting of two Inception V3 networks for skin lesion classification. Experimental results showed that both the segmentation and classification models achieved excellent performances on the ISIC dataset. Lin et al.~\cite{lin2017skin} compared two methods, i.e., U-net and a $C$-Means based approach, for skin lesion segmentation. When evaluated on the ISIC dataset, U-net and $C$-Means based approach achieved $77\%$ and $61\%$ Dice coefficient indices respectively. The results showed that U-net achieved a significantly better performance compared to the clustering method.

Based on the previous two important architectures, a series of deep learning models were developed for skin lesion segmentation. Yuan~\cite{yuan2017automatic} proposed a framework based on deep fully convolutional-deconvolutional neural networks to automatically segment skin lesions in dermoscopy images. The method was tested on the ISIC dataset and took the first place with an average Jaccard index of $0.784$ on the validation dataset. Later, Yuan et al.~\cite{yuan2017improving} extended their previous work~\cite{yuan2017automatic} by proposing a deeper network architecture with smaller kernels to enhance its discriminant capacity. Moreover, color information from multiple color spaces was included to facilitate network training. When evaluated on the ISIC dataset, the method achieved an average Jaccard index of $0.765$, which took the first place in the challenge then. Codella et al.~\cite{codella2017deep} proposed a fully-convolutional U-Net structure with joint RGB and HSV channel inputs for skin lesion segmentation. Experimental results showed that the proposed method obtained competitive segmentation performance to state-of-the-art, and presented agreement with the groundtruth that was within the range of human experts. Al-Masni et al.~\cite{al2018skin} developed a skin lesion segmentation method via deep full resolution convolutional networks. The method was able to directly learn full resolution result of each input image without the need of preprocessing or postprocessing operations. The method achieved an average Jaccard index of $77.11\%$ and overall segmentation accuracy of $94.03\%$ on the ISIC dataset, and $84.79\%$ and $95.08\%$ on the PH2 dataset, respectively. Ji et al.~\cite{ji2018segmentation} proposed a skin image segmentation method based on salient object detection. The proposed method modified the original U-net by adding a hybrid convolution module to skip connections between the down-sampling and up-sampling stages. Besides, the method employed a deeply supervised structure at each stage of up-sampling to learn from the output features and ground truth. Finally, the multi-path outputs were integrated to obtain better performance. Canalini~\cite{canalini2019skin} proposed a novel strategy to perform skin lesion segmentation. They explored multiple pretrained models to initialize a feature extractor without the need of employing biases-inducing datasets. An encoder-decoder segmentation architecture was employed to take advantage of each pretrained feature extractor. In addition, GANs were used to generate both the skin lesion images and corresponding segmentation masks, serving as additional training data. Tschandl et. al.~\cite{tschandl2019domain} trained VGG and ResNet networks on images from the HAM$10000$ dataset~\cite{tschandl2018ham10000} and then transferred corresponding layers as encoders into the LinkNet model~\cite{chaurasia2017linknet}. The model with transferred information was further trained for a binary segmentation task on the official ISIC 2017 challenge dataset~\cite{codella2018skin}. Experimental results showed that the model with fine-tuned weights achieved a higher Jaccard index than that obtained by the network with random initializations on the ISIC 2017 dataset.

Considering the excellent performance of ResNet~\cite{he2016deep} and DenseNet~\cite{huang2017densely} in image classification tasks, people incorporated the idea of residual block or dense block into existing image segmentation architectures to design effective deep networks for skin lesion segmentation. For example, Yu et al.~\cite{yu2016automated} claimed that they were the first to apply very deep CNNs to automated melanoma recognition. They first constructed a fully convolutional residual network (FCRN) which incorporated multi-scale feature representations for skin lesion segmentation. Then the trained FCRN was utilized to extract patches with lesion regions from skin images and the patches were used to train a very deep residual network for melanoma classification. The proposed framework ranked the first in classification competition and the second in segmentation competition on the ISIC dataset. Li et al.~\cite{li2018dense} proposed a dense deconvolutional network for skin lesion segmentation based on residual learning. The network consisted of dense deconvolutional layers, chained residual pooling, and hierarchical supervision. The method can be trained in an end-to-end manner without the need of prior knowledge or complicated postprocessing procedures and obtained $0.866\%$, $0.765\%$, and $0.939\%$, of Dice coefficient, Jaccard index, and accuracy, respectively, on the ISIC dataset. Li et al.~\cite{lihang2018skin} proposed a dense deconvolutional network for skin lesion segmentation based on encoding and decoding modules. The proposed network consisted of convolution units, dense deconvolutional layers (DDL) and chained residual pooling blocks. Specifically, DDL was adopted to restore the original high resolution input via upsampling, while the chained residual pooling was for fusing multi-level features. In addition, hierarchical supervision was enforced to capture low-level detailed boundary information.

Recently, GANs~\cite{goodfellow2014generative} have achieved great success in image generation and image style transfer tasks. The idea of adversarial training was adopted by people for constructing effective semantical segmentation networks and achieved promising results~\cite{luc2016semantic}. In particular, there have been a few works utilizing GANs for skin disease image segmentation~\cite{wei2019attention,jiang2019decision,bi2019improving,tu2019segmentation}. Udrea et al.~\cite{udrea2017generative} proposed a deep network based on GANs for segmentation of both pigmented and skin colored lesions in images acquired with mobile devices. The network was trained and tested on a large set of images acquired with a smart phone camera and achieved a segmentation accuracy of $91.4\%$. Peng et al.~\cite{peng2019segmentation} presented a segmentation architecture based on adversarial networks. Specifically, the architecture employed a segmentation network based on U-net as generator and a network consisting of certain number of convolutional layers as discriminator. The method was tested on the PH$2$ and ISIC datasets, achieving an average segmentation accuracy of $0.97$ and dice coefficient of $0.94$. Sarker et al.~\cite{sarker2019mobilegan} proposed a lightweight and efficient GAN model (called MobileGAN) for skin lesion segmentation. The MobileGAN combined $1$-D non-bottleneck factorization networks with position and channel attention modules in a GAN model. With only $2.35$ million parameters, the MobileGAN still obtained comparable performance with an accuracy of $97.61\%$ on the ISIC dataset. Singh et al.~\cite{singh2019fca} presented a skin lesion segmentation method based on a modified conditional GAN (cGAN). They introduced a new block (called factorized channel attention, FCA) into the encoder of cGAN, which exploited both channel attention mechanism and residual $1$-D kernel factorized convolution. In addition, multi-scale input strategy was utilized to encourage the development of filters that were scale-variant.

Besides designing novel architectures, people also considered developing effective deep learning models for skin lesion segmentation from other aspects. For example, Jafari et al.~\cite{jafari2016skin} proposed a deep CNN architecture to segment the lesion regions of skin images taken by digital cameras. Local and global patches were utilized simultaneously such that the CNN architecture was able to capture the global and local information of images. Experimental results on the Dermquest dataset showed that the proposed method obtained a high accuracy of $98.5\%$ and sensitivity of $95.0\%$. Yuan et al.~\cite{yuan2017automatic} proposed a new loss function for a deep network to adapt it to a skin lesion segmentation task. Specifically, they designed a novel loss function based on the Jaccard distance for a fully convolutional neural network and performed skin lesion segmentation on dermoscopy images. CNNs for skin lesion segmentation commonly accept low-resolution images as inputs to reduce computational cost and network parameters. This situation may lead to the loss of important information contained in images. To resolve this issue and develop a resolution independent method for skin lesion segmentation, {\"U}nver et al.~\cite{unver2019skin} proposed a method by combining the YOLO model and GrabCut algorithm for skin lesion segmentation. Specifically, the YOLO model was first employed to locate the lesions and image patches were extracted according to the location results. Then the GrabCut algorithm was utilized to perform segmentation on the image patches. Due to the small size of the labeled training dataset and large variations of skin lesions, the generalization property of segmentation models is limited. To address this issue, Cui et al.~\cite{cui2019ensemble} proposed an ensemble transductive learning strategy for skin lesion segmentation. By learning directly from both training and testing sets, the proposed method can effectively reduce the subject-level difference between training and testing sets. Thus, the generalization performance of existing segmentation models can be improved. Soudani et al.~\cite{soudani2019image} proposed a segmentation method based on crowdsourcing and transfer learning for skin lesion extraction. Specifically, they utilized two pretrained networks, i.e., VGG-16 and ResNet-50, to extract features from the convolutional parts. Then a classifier with an output layer composed of five nodes was built. In this way, the proposed method was able to dynamically predict the most appropriate segmentation technique for the detection of skin lesions in any input image.

For convenient reference, we list the aforementioned works on skin lesion segmentation with deep learning methods in Table~\ref{seg1} and Table~\ref{seg2}.

\begin{table}
\caption{References of skin lesion segmentation with deep learning (part $1$).}\label{seg1}
\centering
\scriptsize
\begin{tabular}{p{1.0cm}<{\centering}cp{1.8cm}<{\centering}p{2.0cm}<{\centering}p{4.6cm}}
\hline
{\bf Reference}                &  {\bf Year}     & {\bf Dataset}                & {\bf No. of images}   & {\bf Segmentation method}         \\
\hline
\cite{jafari2016skin}          &  2016           & Derm101                      & 126                   & A CNN architecture consisting of two subpaths, with one accounting for global information and another for local information.\\
\cite{yu2016automated}         &  2016           & ISIC                         & 1,250                 & Fully convolutional residual network.\\
\cite{bi2017dermoscopic}       &  2017           & ISIC and PH2                 & 1,279 and 200         & Multistage fully convolutional networks with parallel integration.\\
\cite{goyal2017multi}          &  2017           & ISIC                         & 2,750                 & A transfer learning approach which uses both
partial transfer learning and full transfer learning to train FCNs for multi-class semantic segmentation.\\
\cite{lin2017skin}             &  2017           & ISIC                         & 2,000                 & U-Nets with a histogram equalization based preprocessing step.\\
\cite{codella2017deep}         &  2017           & ISIC                         & 1,279                 &  An ensemble system combining traditional machine learning methods with deep learning methods. \\
\cite{attia2017skin}           &  2017           & ISIC                         & 1,275                 & An architecture combining an auto-encoder network with a four-layer recurrent network with four decoupled directions.  \\
\cite{chang2017skin}           &  2017           & ISIC                         & 2,000                 & A deep network similar as U-net. \\
\cite{yuan2017automatic}       &  2017           & ISIC and PH2                 & 1,279 and 200         & A fully convolutional neural network with a novel loss function defined based on the Jaccard distance. \\
\cite{yuan2017improving}       &  2017           & ISIC                         & 2,750                 & A convolutional-deconvolutional neural network.\\
\cite{udrea2017generative}     &  2017           & A proprietary database       & 3,000                 & A GAN with U-net being the generator.\\
\cite{al2018skin}              &  2018           & ISIC and PH2                 & 2,750 and 200         & A full resolution convolutional network.\\
\cite{ji2018segmentation}      &  2018           & From ISIC and other sources  & 2,600                 & Modified U-net with hybrid convolution modules and deeply supervised structure.\\
\cite{li2018dense}             &  2018           & ISIC                         & 2,900                 & A dense deconvolutional network based on residual learning.\\
\cite{lihang2018skin}          &  2018           & ISIC                         & 1,950                 & A dense deconvolutional network based on encoding and decoding modules.\\
\cite{canalini2019skin}        &  2019           & ISIC                         & 10,015                & An encoder-decoder architecture with multiple pretrained models as feature extractors. In addition, GANs were used to generate additional training data.\\
%\cite{unver2019skin}           &  2019           & ISIC and PH2                 & 2750 and 200          & Detect the skin lesion location with YOLO model and segment the image with GrabCut algorithm.\\
%\cite{tschandl2019domain}      &  2019           & HAM10000, ISIC and PH2       & Around 20,000         & A LinkNet architecture with pretrained ResNet as encoders.\\
%\cite{phillips2019segmentation}&  2019           & \\
%\cite{peng2019segmentation}    &  2019\\
%\cite{sarker2019mobilegan}     &  2019\\
%\cite{singh2019fca}            &  2019\\
%\cite{cui2019ensemble}         &  2019\\
%\cite{soudani2019image}        &  2019\\
\hline
\end{tabular}
\end{table}

\begin{table}
\caption{References of skin lesion segmentation with deep learning (part $2$).}\label{seg2}
\centering
\scriptsize
\begin{tabular}{p{1.0cm}<{\centering}cp{1.8cm}<{\centering}p{2.0cm}<{\centering}p{4.6cm}}
\hline
{\bf Reference}                &  {\bf Year}     & {\bf Dataset}                & {\bf No. of images}   & {\bf Segmentation method}         \\
\hline
\cite{unver2019skin}           &  2019           & ISIC and PH2                 & 2750 and 200          & Detect skin lesion location with the YOLO model and segment images with the GrabCut algorithm.\\
\cite{tschandl2019domain}      &  2019           & HAM10000, ISIC and PH2       & Around 20,000         & A LinkNet architecture with pretrained ResNet as encoders.\\
\cite{phillips2019segmentation}&  2019           & TCGA                         & $50$                  & A multi-stride fully convolutional network. \\
\cite{peng2019segmentation}    &  2019           & ISIC and PH$2$               & $1,279$ and $200$     & An architecture based on adversarial networks with a segmentation network based on U-net and a discrimination network linked by certain convolutional layers.\\
\cite{sarker2019mobilegan}     &  2019           & ISIC                         & $3,344$               & MobileGAN combining $1$-D non-bottleneck factorization networks with position and channel attention modules. \\
\cite{singh2019fca}            &  2019           & ISBI 2016, ISBI 2017 and ISIC & $1,279$, $2,750$ and $3,694$ & A modified cGAN with factorized channel attention as the encoder.\\
\cite{cui2019ensemble}         &  2019           & ISIC                         & $3,694$               & A transductive approach which chooses some of the pixels in test images to participate the training of the segmentation model together with the training set.\\
\cite{soudani2019image}        &  2019           & ISIC                         & $2,750$               & A segmentation recommender based on crowdsourcing and transfer learning.\\
\hline
\end{tabular}
\end{table}

\subsubsection{Skin disease classification}

\begin{figure}
\center
\includegraphics[width=0.8\textwidth]{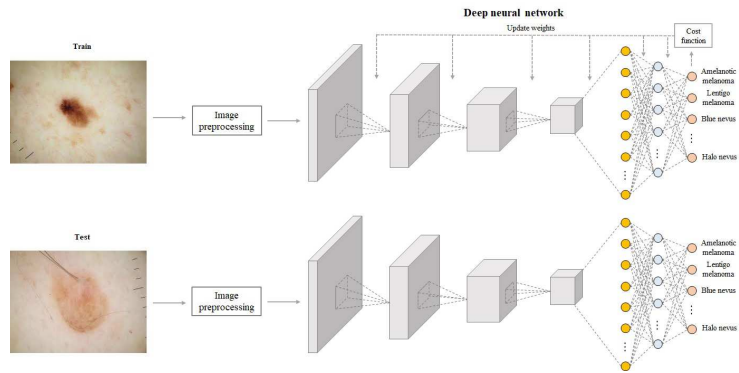}
\caption{The workflow for a typical skin disease classification task.} \label{workflow_class}
\end{figure}

Skin disease classification is the last step in the typical workflow of a CAD system for skin disease diagnosis. Depending on the purpose of the system, the output of a skin disease classification algorithm can be binary (e.g., benign and malignant), ternary (e.g., melanoma, dysplastic nevus and common nevus) or $n\geq4$ categories. To accomplish the task of classification, various deep learning methods have been proposed to classify skin disease images. In the following, we present a brief review of the existing deep learning methods for skin disease classification. The workflow for a typical skin disease classification task is illustrated in Fig.~\ref{workflow_class}.

Initially, traditional machine learning methods were employed to extract features from skin images and then the features were input to a deep learning based classifier for classification. The study by Masood et al.~\cite{masood2015self} was one of the earliest works that applied modern deep learning methods to skin disease classification tasks. The authors first detected skin lesions with a histogram based thresholding algorithm, and then extracted features with three machine learning algorithms. Finally, they classified the features with a semi-supervised classification model that combined DBNs and a self-advising support vector machine (SVM)~\cite{maali2013self}. The proposed model was tested on a collection of $100$ dermoscopy images and achieved better results than other popular algorithms. Premaladha et al.~\cite{premaladha2016novel} proposed a CAD system to classify dermoscopy images of melanoma. With enhanced images, the system segmented affected skin lesion from normal skin. Then fifteen features were extracted from these segmented images with a few machine learning algorithms and input to a deep neural network for classification. The proposed method achieved a classification accuracy of $93\%$ on the testing data.

With the development of deep learning, more and more novel networks are designed such that they can be trained in an end-to-end manner. In particular, various such kind of advanced deep networks were proposed for skin disease classification in the past few years. In 2016, Nasr et al.~\cite{nasr2016melanoma} implemented a CNN for melanoma classification with non-dermoscopy images taken by digital cameras. The algorithm can be applicable in web-based and mobile applications as a telemedicine tool and also as a supporting system to assist physicians. Demyanov et al.~\cite{demyanov2016classification} trained a five-layer CNN for classifying two types of skin lesion data. The method was tested on the ISIC dataset and the best mean classification accuracies for the ``Typical Network" and ``Regular Globules" datasets were $88\%$ and $83\%$, respectively. In 2017, Esteva et al.~\cite{esteva2017dermatologist} trained a single CNN using only pixels and disease labels as inputs for skin lesion classification. The dataset in their study consists of $129,450$ clinical images of $2,032$ different diseases. Moreover, they compared the performance of the CNN with $21$ board-certified dermatologists on biopsy-proven clinical images with two critical binary classification use cases: keratinocyte carcinomas versus benign seborrheic keratoses; and malignant melanomas versus benign nevi. Results showed that the CNN achieved performances on par with all tested experts across both tasks, demonstrating that an artificial intelligence was capable of classifying skin cancer with a level of competence comparable to dermatologists. Walker et. al.~\cite{walker2019dermoscopy} reported a work on dermoscopy images classification which evaluated two different inputs derived from a dermoscopy image: visual features determined via a deep neural network (System A) based on the Inception V2 network~\cite{ioffe2015batch}; and sonification of deep learning node activations followed by human or machine classification (System B). A laboratory study (LABS) and a prospective observational study (OBS) each confirmed the accuracy level of this decision support system. In both LABS and OBS, System A was highly specific and System B was highly sensitive. Combination of the two systems potentially facilitated clinical diagnosis. Brinker et. al.~\cite{brinker2019convolutional} trained a CNN with dermoscopy images from the HAM$10000$ dataset exclusively for identifying melanoma in clinical photographs. They compared the performance of the automated digital melanoma classification algorithm with that of $145$ dermatologists from $12$ German university hospitals. This was the first time that a CNN without being trained on clinical images performed on par with dermatologists on a clinical image classification task.

Generally, deep neural networks have a high variance and it can be frustrating when trying to develop a final model for decision making. One solution to this issue is to train multiple models instead of a single one and combine the predictions from these models to form the final results, which is called ensemble learning~\cite{singh2016swapout,lee2017ensemble}. Ensemble learning commonly produces better results than any of the single model, and has been applied to skin disease classification. Han et al.~\cite{han2018deep} created datasets of standardized nail images using a region-based CNN (R-CNN). Then the datasets were utilized to fine-tune the pretrained ResNet-152 and VGG-19 networks. The outputs of the two networks were combined together and input to a two-hidden-layered feedforward neural network for final prediction. Experimental results showed that the diagnostic accuracy for onychomycosis using deep learning was superior to that of most of the dermatologists who participated in this study. Though CNNs achieved expert-level accuracy in the diagnosis of pigmented melanocytic lesions, the most common types of skin cancer are nonpigmented and nonmelanocytic which are difficult to be diagnosed. Tschandl et al.~\cite{tschandl2019expert} trained a model combining the Inception V$3$ and ResNet-50 for skin lesion classification with $7,895$ dermoscopy and $5,829$ close-up images and tested the model on a set of $2,072$ images. The authors compared the performance of the model with $95$ human raters and the results showed that the model can classify dermoscopy and close-up images of nonpigmented lesions as accurate as human experts in the experimental settings. Mahbod et al.~\cite{mahbod2019fusing} proposed a hybrid CNN ensemble scheme that combined intra-architecture and inter-architecture networks for skin lesion classification. Through fine-tuning networks of different architectures with different settings and combining the results from multiple sets of fine-tuned networks, the proposed method yielded excellent results in the ISIC 2017 skin lesion classification challenge without requiring extensive preprocessing, or segmentation of lesion areas, or additional training data. Perez et al.~\cite{perez2019solo} evaluated $9$ different CNN architectures for melanoma classification, with $5$ sets of splits created on the ISIC Challenge 2017 dataset, and $3$ repeated measures, resulting in $135$ models. The author found that ensembles of multiple models can always outperform the individual model.

Despite deep learning models achieved excellent performance on various experimental datasets, one should also consider the fact that most deep learning models require a whole lot of labeled data for training, and obtaining vast amounts of labeled data (especially medical data) can be really difficult and expensive in terms of both time and money. Fortunately, transfer learning~\cite{yosinski2014transferable,shin2016deep} can be a strategy to alleviate this issue, enabling deep learning models to achieve satisfying performance on small datasets. The basic concept of transfer learning is to train a model on a large dataset and transfer its knowledge to a smaller one. Thus, one can utilize a deep network trained on unrelated categories in a massive dataset (usually ImageNet~\cite{deng2009imagenet}) and apply it to our own problems (e.g., skin disease classification).

As only limited skin disease data can be obtained publicly, transfer learning is widely adopted in skin disease classification tasks. Liao~\cite{liao2016deep} used the pretrained VGG-$16$, VGG-$19$ and GoogLeNet networks to construct a universal skin disease diagnosis system. The author trained the networks on the Dermnet dataset and tested them on the Dermnet and OLE datasets. When tested on the Dermnet dataset, the proposed system achieved a top-$1$ accuracy of $73.1\%$ and top-$5$ accuracy of $91\%$, respectively. For the test on the OLE dataset, the top-$1$ and top-$5$ accuracies are $31.1\%$ and $69\%$, respectively. In a more recent work by Liao et al.~\cite{liao2016skin}, the authors utilized the pretrained AlexNet for both disease-targeted and lesion-targeted classification tasks. They pointed out that lesion type tags should also be considered as the target of an automated diagnosis system such that the system can achieve a high accuracy in describing skin lesions. Kawahara et al.~\cite{kawahara2016deep} extracted multi-scale features of skin lesions with a pretrained fully convolutional AlexNet. Then the features were pooled and used to train a logistic regression classifier to classify non-dermoscopic skin images. Sun et al.~\cite{sun2016benchmark} built a benchmark dataset for clinical skin diseases and fine-tuned the pretrained VGG-16 model on the dataset for skin disease classification. Zhang et al.~\cite{zhang2017computer} utilized the pretrained Inception V$3$ network to classify dermoscopy images into four classes. The model was evaluated on a private dataset collected from the Peking Union Medical College Hospital and experimental results showed that deep learning algorithms were promising for automated skin disease diagnosis. Fujisawa et al.~\cite{fujisawa2019deep} proposed to apply the pretrained GoogleLetNet to skin tumor classification with a dataset containing $4,867$ clinical images of $21$ skin diseases. Compared with board-certified dermatologists, the algorithm achieved better performances with an accuracy of $92.4\%\pm2.1\%$. Lopez et al.~\cite{lopez2017skin} utilized the VGG-16 network to perform melanoma classification. The authors trained the network in three different ways: 1) training the network from scratch; 2) using the transfer learning paradigm to leverage features from a VGG-net pretrained on ImageNet; and 3) performing the transfer learning paradigm and fine-tuning the network. In the experiments, the proposed approach achieved state-of-the-art classification results with a sensitivity of $78.66\%$ and precision of $79.74\%$. Han et. al.~\cite{han2018classification} employed the pretrained ResNet-152 model to classify clinical images of $12$ skin diseases. The model was further fine-tuned with $19,398$ images from multiple dermoscopy image datasets. Haenssle et al.~\cite{haenssle2018man} employed a pretrained Inception V4 network for melanoma classification. In the study, the authors compared the performance of the algorithm with that of an international group of $58$ dermatologists. The results demonstrated that the performance of CNN outperformed that of most but not all dermatologists. Zhang et al.~\cite{zhang2018towards} utilized the Inception V$3$ network to classify dermoscopy images of four common skin diseases. To further facilitate the application of the algorithm to CAD support, the authors generated a hierarchical semantic structure based on domain expert knowledge to represent classification/diagnosis scenarios. The proposed algorithm achieved an accuracy of $87.25\pm2.24\%$ on the testing dataset. Joanna et al.~\cite{jaworek2019melanoma} proposed to perform preoperative melanoma thickness evaluation with a pretrained VGG-$19$ network. Experimental results showed that the developed algorithm achieved state-of-the-art melanoma thickness prediction result with an overall accuracy of $87.2\%$. Yu et al.~\cite{yu2018melanoma} proposed a novel framework for dermoscopy image classification. Specifically, the authors first extracted image representations via a pretrained deep residual network and obtained global image descriptors with the fisher vector encoding method. After that, the obtained descriptors were utilized to classify melanoma images with SVM. Menegola et al.~\cite{menegola2017knowledge} systematically investigated the applications of knowledge transfer of deep learning in dermoscopy image recognition. Their results suggested that transfer learning from a related task can lead to better results on target tasks. Hekler et al.~\cite{hekler2019pathologist} claimed that they were the first to implement a deep learning method for histopathologic melanoma diagnosis and compare the performance of the algorithm with that of an experienced histopathologist. In the study, they utilized a pretrained ResNet-50 network~\cite{he2016deep} to classify histopathologic slides of skin lesions into classes of nevi and melanoma. They demonstrated that the discordance between the CNN and expert pathologist was comparable with that between different pathologists as reported in the literature. Polevaya et al.~\cite{polevaya2019skin} utilized the pretrained VGG-16 network to classify primary morphology images of macule, nodule, papule and plaque. Experimental results showed that the method was able to achieve an accuracy of $77.50\%$ for $4$ classes and $81.67\%$ for $3$ classes on the testing dataset.

Attention mechanism aims to learn a context vector to weight the input such that salient features can be highlighted and unrelated ones can be suppressed. It was first extensively used in the field of natural language processing (NLP)~\cite{yang2016stacked,liu2016learning}, and has been applied to skin disease classification recently. Barata et al.~\cite{barata2019deep} proposed a hierarchical attention model combining CNNs with LSTM and attention modules for skin disease classification. The model made use of the hierarchical organization of skin lesions, as identified by dermatologists, so as to incorporate medical knowledge into the decision process. Particularly, the attention modules were able to identify relevant regions in the skin lesions and guide the classification decision. The proposed approach achieved state-of-the-art results on the two dermoscopy datasets of ISIC 2017 and ISIC 2018.

As discussed above, GANs~\cite{goodfellow2014generative} with the capability of generating synthetic real-world like samples developed rapidly during the past few years. In particular, GANs or the ideas of adversarial training have been utilized to construct effective algorithms for skin diseases classification~\cite{rashid2019skin}. The applicability of deep learning methods to melanoma detection is compromised by the limitation of available skin lesion datasets that are small, heavily imbalanced, and contain images with occlusions. To alleviate this issue, Bisla et al.~\cite{bisla2019towards} proposed to purify data with deep learning based methods and augment data with GANs, for populating scarce lesion classes, or equivalently creating virtual patients with predefined types of lesions. These preprocesses can be used in a deep neural network for lesion classification. Experimental results showed that the proposed preprocesses can boost the performance of a deep neural network in melanoma detection. Yi et al.~\cite{yi2018unsupervised} utilized the categorical GAN assisted by Wasserstein distance for dermoscopy image classification in an unsupervised and semi-supervised way. Experimental results on the ISIC dataset showed that the proposed method achieved an average precision score of $0.424$ with only $140$ labeled images. In addition, the method was able to generate real-world like dermoscopy images. Gu et al.~\cite{gu2019progressive} proposed two methods for cross-domain skin disease classification. They first explored a two-step progressive transfer learning technique by fine-tuning pretrained networks on two skin disease datasets. Then they utilized adversarial learning as a domain adaptation technique to perform invariant attribute translation from source domain to target domain. Evaluation results on two skin disease datasets showed that the proposed method was effective in solving the domain shift problem.

Besides the above research directions for skin disease classification, people also worked on the problem of skin disease classification from other aspects. Mishra et al.~\cite{mishra2019interpreting} investigated the effectiveness of current deep learning methods for skin disease classification. The authors analyzed the classification processes of several deep neural networks (including Resnet-$34$, ResNet-$50$, ResNet-$101$ and ResNet-$152$) for common East Asian dermatological conditions. The authors chose ten common categories of skin diseases based on their prevalence for evaluation. With an accuracy of more than $85\%$ in the experiments, the authors tried to investigate why existing models were unable to achieve comparable results with those in object identification tasks. The study suggested that the deep learning based dermoscopy identification and dataset creation can be improved. By integrating segmentation results with skin disease classification process, better classification results tend to be obtained. Wan~\cite{wan2018deep} implemented several deep networks (including U-net, Deeplab, Inception V3, MobileNet~\cite{howard2017mobilenets} and NASNet~\cite{esteva2017dermatologist}) for skin lesion segmentation and classification on the ISIC 2017 challenge dataset. Particularly, the author cropped skin images with the trained segmentation model and trained a classification model based on the cropped data. In this way, the classification accuracy was further improved. Shi et al.~\cite{shi2019active} presented a novel active learning framework for cost-effective skin lesion analysis. They proposed a dual-criteria to select samples and an intraclass sample aggregation scheme to enhance the model. Using only up to $50\%$ of samples, the proposed approach achieved state-of-the-art performance on both tasks on the ISIC dataset. Tschandl et al.~\cite{tschandl2019diagnostic} trained a neural network to classify dermatoscopy images from three retrospectively collected image datasets. The authors obtained diagnosis predictions through two ways, i.e., based on the most commonly occurring diagnosis in visually similar images (obtained via content-based image retrieval), or based on the top-$1$ class prediction of the network. Experimental results showed that presenting visually similar images based on features from a network showed comparable accuracy with the softmax probability-based diagnoses of deep networks.

For convenient comparison, we list the references of skin disease image classification with deep learning in Table~\ref{classification} and~\ref{classification2}.

\begin{table}
\caption{References of skin disease classification with deep learning (part 1).}\label{classification}
\centering
\scriptsize
\begin{tabular}{p{1.0cm}<{\centering}p{0.3cm}<{\centering}p{3.5cm}<{\centering}p{1.0cm}<{\centering}p{4.8cm}}
\hline
{\bf Reference}                      &  {\bf Year}     & {\bf Dataset}             & {\bf No. of data}     & {\bf Classification method}         \\
\hline
\cite{masood2015self}                &  2015           & Self-collected dataset    & 290                   & Detect skin lesions with a thresholding algorithm, extract features with three machine learning algorithms, and perform classification with a model combining DBNs and self-advised SVM.\\
\cite{demyanov2016classification}    &  2016           & ISIC                      & 29,323                & A CNN with three convolutional layers and pooling layers, and two fully-connected layers.\\
\cite{liao2016deep}                  &  2016           & Dermnet and OLE           & $>$24,300             & Pretrained VGG-16, VGG-19 and GoogLeNet.     \\
\cite{liao2016skin}                  &  2016           & Self-collected dataset    & 75,665                & Pretrained AlexNet. \\
\cite{kawahara2016deep}              &  2016           & Dermofit Image Library    & 1,300                 & Pretrained fully convolutional AlexNet.\\
\cite{nasr2016melanoma}              &  2016           & MED-NODE                  & 170                   & A CNN with two convolutional layers and two fully-connected layers. \\
\cite{sun2016benchmark}              &  2016           & SD-198                    & 6,584                 & Pretrained VGG-16.\\
\cite{premaladha2016novel}           &  2016           & Self-collected dataset    & 992                   & Segment skin lesions with Otsu's method and extract fifteen features with several algorithms, then classify images with deep networks and a hybrid adaboost-SVM. \\
\cite{zhang2017computer}             &  2017           & Collected from the Peking Union Medical College Hospital & $>$28,000 & Pretrained GoogleNet Inception V3.\\
\cite{esteva2017dermatologist}       &  2017           & Images from $18$ online repositories and clinical data from the Stanford University Medical Center. & 129,450 & Pretrained GoogleNet Inception V3. \\
\cite{fujisawa2019deep}              &  2017           & Images collected from the University of Tsukuba Hospital & 4,867 & Pretrained GoogLeNet.\\
\cite{lopez2017skin}                 &  2017           & ISIC                      & 1,279                 & VGG-16 trained from scatch and pretrained VGG-16.\\
\cite{menegola2017knowledge}         &  2017           & IAD and ISIC              & $\geq$1000 and 1,279  & Transfer learning with VGG-M model.\\
\cite{han2018classification}         &  2018           & Asan dataset, MED-NODE dataset, atlas site images, Hallym and Edinburgh datasets  & 19,878 & Pretrained ResNet-152. \\
\cite{han2018deep}                   &  2018           & Self-collected dataset    & 54,666                & The outputs of the pretrained ResNet-152 and VGG-19 are combined together and input to two fully-connected layers for classification. \\
\hline
\end{tabular}
\end{table}

\begin{table}
\caption{References of skin disease classification with deep learning (part 2).}\label{classification2}
\centering
\scriptsize
\begin{tabular}{p{1.0cm}<{\centering}p{0.3cm}<{\centering}p{3.5cm}<{\centering}p{1.0cm}<{\centering}p{4.8cm}}
\hline
{\bf Reference}                      &  {\bf Year}     & {\bf Dataset}             & {\bf No. of data}     & {\bf Classification method}         \\
\hline
\cite{haenssle2018man}               &  2018           & Test-set-300 and ISIC     & 300 and 100           & Pretrained GoogLeNet Inception V4.\\
\cite{zhang2018towards}              &  2018           & Collected from the Peking Union Medical College Hospital & $>$2,800 & Pretrained GoogleNet Inception V3.\\
\cite{yu2018melanoma}                &  2018           & ISIC                      & 1,279                 & Extract image features via
a pretrained CNN, obtain global descriptors based on fisher vector encoding method and perform classification with SVM.\\
\cite{wan2018deep}                   &  2018           & ISIC                      & $>$ 20,000            & GoogLeNet Inception V3, MobileNet and NASNet.\\
\cite{yi2018unsupervised}            &  2018           & ISIC and PH2              & 1,279 and 200         & Categorical GAN assisted by Wasserstein distance.\\
\cite{perez2019solo}                 &  2019           & ISIC                      & 2,750                 & 9 different pretrained networks. \\
\cite{barata2019deep}                &  2019           & ISIC                      & 2,750                 & A model combining CNNs with LSTM and attention modules.\\
\cite{jaworek2019melanoma}           &  2019           & IAD                       & 244                   & Pretrained VGG-19. \\
\cite{tschandl2019expert}            &  2019           & Self-collected dataset    & 15,796                & A model combining GoogLeNet Inception V3 and ResNet50. \\
\cite{brinker2019convolutional}      &  2019           & ISIC and HAM10000         & 20,735                & Pretrained ResNet50.    \\
\cite{walker2019dermoscopy}          &  2019           & ISIC and IAD              & 2,361 and 2,800       & GoogLeNet Inception V2. \\
\cite{mahbod2019fusing}              &  2019           & ISIC                      & 2,787                 & Combine multiple networks (AlexNet, VGG-nets and ResNet), and fine-tune the networks multiple times and ensemble the multiple results. \\
\cite{tschandl2019diagnostic}        &  2019           & EDRA, ISIC and PRIV       & 888, 2,750 and 16,691 & Pretrained ResNet-50. \\
\cite{mishra2019interpreting}        &  2019           & Self-collected            & 7,264                 & Pretrained Resnet-$34$, ResNet-$50$, ResNet-$101$, and ResNet-$152$.\\
\cite{hekler2019pathologist}         &  2019           & Collected from the institute of Dr. Krahl   & 695 & Pretrained ResNet-50.\\
\cite{bisla2019towards}              &  2019           & ISIC, PH2 and Dermofit Image Library    & 3,982   & Segment lesions with U-net, generate data with DCGANs~\cite{radford2015unsupervised} and classify lesions with pretrained ResNet-50.\\
\cite{polevaya2019skin}              &  2019           & Self-collected            & -                     & Pretrained VGG-16.\\
\cite{gu2019progressive}             &  2019           & MoleMap and HAM1000           & 102,451 and 10,015 & Progressive transfer learning of deep CNN models and GAN based method. \\
\cite{shi2019active}                 &  2019           & ISIC                      & 3,582                 & A novel active learning method.\\
\hline
\end{tabular}
\end{table}

\subsubsection{Multi-task learning for skin disease diagnosis}

\begin{figure}
\center
\includegraphics[width=0.6\textwidth]{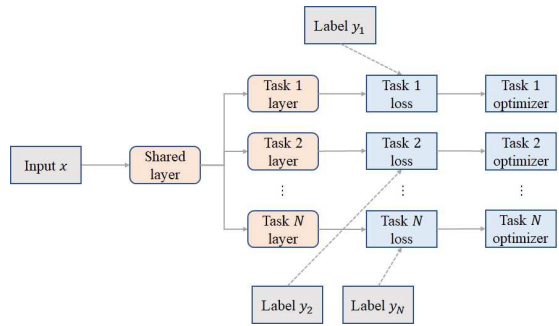}
\caption{The workflow for a typical multi-task learning.} \label{multi-task}
\end{figure}

In machine learning, people generally train a single model or an ensemble of models to complete their desired tasks. While they can achieve acceptable results in this way, information that might contribute to better performance is ignored. Specifically, this information comes from the training data of related tasks. By sharing representations among related tasks, existing models are able to generalize better in the original task. This approach is called multi-task learning (MTL)~\cite{ruder2017overview}. MTL enables multiple learning tasks to be solved simultaneously, while exploring the commonalities and differences across tasks. This can result in the improvement of learning efficiency and prediction accuracy of the task-specific models, when compared to training models separately~\cite{caruana1997multitask}. The workflow for a typical MTL is illustrated in Fig.~\ref{multi-task}.

Many works have adopted MTL for skin disease diagnosis. Yang et. al.~\cite{yang2017novel} proposed a multi-task CNN based model for skin lesion analysis. In the model, each input dermoscopy image is associated with multiple labels that describe different characteristics of the skin lesion. Then multi-task methods are utilized to perform skin lesion segmentation and classifications simultaneously. Experimental results showed that the multi-task method achieved promising performance in both tasks. Different with existing deep learning approaches that commonly use two networks to separately perform lesion segmentation and classification, Li et. al.~\cite{li2018skin} proposed a deep learning framework consisting of multi-scale fully convolutional residual networks and a lesion index calculation unit to simultaneously perform the two tasks. To investigate the correlation between skin lesions and their body site distributions, the authors in work~\cite{haofu2017deep} trained a deep multi-task learning framework to jointly optimize skin lesion classification and body location classification. The experimental results verified that features jointly learned with body location information indeed boosted the performance of skin lesion classification. Kawahara et al.~\cite{kawahara2018seven} proposed a multi-task deep neural network, trained on a multi-modal dataset (including clinical and dermoscopy images, and patient meta-data), to classify the $7$-point melanoma checklist criteria and perform skin lesion diagnosis. The network trained with several multi-task loss functions was able to handle the combination of input modalities. The model classified the $7$-point checklist and performed skin condition diagnosis, and produced multi-modal feature vectors suitable for image retrieval and localization of clinically discriminative regions.

\subsubsection{Miscellany}

Apart from the above applications of deep learning in skin disease diagnosis, there are several works applying deep learning to skin disease diagnosis from other aspects.

GANs have been utilized to synthesize skin images so as to facilitate skin disease diagnosis~\cite{ghorbani2019dermgan,ali2019data,nyiri2019style}. To address the problems caused by lack of sufficient labeled data in skin disease diagnosis tasks, Bissoto et al.~\cite{bissoto2018skin} proposed to use GAN to generate realistic synthetic skin lesion images. Experimental results showed that they could generate high-resolution (up to $1024\times512$) samples containing fine-grained details. Moreover, they employed a classification network to evaluate the generated images and results showed that the synthetic images comprised clinically meaningful information. With the help of progressive growing and GANs, Baur et al.~\cite{baur2018generating} generated extremely realistic high-resolution dermoscopy images. Experimental results showed that even expert dermatologists found it hard to distinguish the synthetic images from real ones. Therefore, this method can be served as a new direction to deal with the problem of data scarcity and class imbalance. Yang et al.~\cite{yang2019dual} proposed a novel generative model based on a dual discrimination training algorithm for autoencoders to synthesize dermoscopy images. In contrast to other related methods, an adversarial loss was added to the pixel-wise loss during the image construction phase. Through experiments, they demonstrated that the method can be applied to various tasks including data augmentation and image denoising. Baur et al.~\cite{baur2018melanogans} utilized GANs to generate realistically looking high-resolution skin lesion images with only a small training dataset ($2,000$ samples). They both quantitatively and qualitatively compared state-of-the-art GAN architectures such as DCGAN and LAPGAN against a modification of the latter one for the image generation task at a resolution of $256\times256$. Experimental results showed that all the models can approximate the real data distribution. However, major differences when visually rating sample realism, diversity and artifacts can be observed. %By employing the synthetic images for skin lesion classification, they further showed that the problem of heavy class imbalance can be alleviated with the help of synthesized high-resolution melanoma samples.

Besides the above GAN-based applciations, Han et al.~\cite{han2019keratinocytic} proposed a method based on R-CNN for detecting keratinocytic skin cancer on the face. They first used R-CNN to create $924,538$ possible lesions by extracting nodular benign lesions from $182,348$ clinical photographs. After labeling these possible lesions, CNNs were trained with $1,106,886$ image crops to locate and diagnose cancer. Experimental results showed that the proposed algorithm achieved a higher F$1$ score of $0.831$ and Youden index score of 0.675 than those of nondermatologic physicians. Additionally, the accuracy of the algorithm was comparable with that of dermatologists. Galdran et al.~\cite{galdran2017data} utilized computational color constancy techniques to construct an artificial data augmentation method suitable for dermoscopy images. Specifically, they applied the shades of gray color constancy technique to color-normalize images of the entire training set, while retaining the estimated illuminants. Then they drew one sample from the distribution of training set and applied it to the normalized image. They performed experiments on the ISIC dataset by employing this technique to train two CNNs for skin lesion segmentation and classification. Attia et al.~\cite{attia2019digital} proposed a deep learning method based on a hybrid network consisting of convolutional and recurrent layers for hair segmentation with weakly labeled data. Deep encoded features were utilized for detection and delineation of hair in skin images. The encoded features were then fed into the recurrent layers to encode the spatial dependencies between disjointed patches. Experiments conducted on the ISIC dataset showed that the proposed method obtained excellent segmentation results with a Jaccard Index of $77.8\%$ and tumour disturb pattern of $14\%$.

\section{Discussion}

Skin disease diagnosis with deep learning methods has attracted much attention and achieved promising progress in recent years~\cite{esteva2017dermatologist,liao2016deep,lopez2017skin}. In the published literature, the performances achieved by deep learning methods for skin disease diagnosis are similar as those achieved by dermatologists. To develop and validate excellent algorithms or systems supporting new imaging techniques, lots of research and innovative system development are required~\cite{pathan2018techniques}. The major drawback of dermoscopy examination by dermatologists is that the process is subjective and results may vary with experience. Thus, biopsy is needed to differentiate benign cases from malignant ones. Biopsying benign lesions of skin diseases may lead to increased anxieties to patients and aggravate the expense to healthcare systems. Factors, such as training, time, and experience needed to properly utilize various available and upcoming techniques, present a huge barrier to early and accurate diagnosis of skin diseases. Although many automated skin disease diagnosis methods have been developed, a complete decision support system has not been developed.

In this section, we discuss the major challenges faced in the field of skin disease diagnosis with deep learning. Instead of describing specific cases encountered, we focus more on the fundamental challenges and explain the root causes of these issues. Then, we try to provide suggestions to deal with these problems.

\subsection{Challenges}

With the development of deep learning in the past few years, a variety of works on skin disease diagnosis with deep learning methods have been proposed and achieved promising performance. However, there are still several issues that should be resolved before deep learning can be extensively applied to real-life clinical scenarios of skin disease diagnosis.

\subsubsection{Limited labeled skin disease data}
Previous works on skin disease diagnosis with deep learning were commonly trained and tested on datasets with limited number of images. The biggest publicly available skin disease dataset that can be found in literate until now is the ISIC dataset~\cite{codella2018skin} containing more than $20,000$ skin images. Though one may obtain large numbers of skin disease data without any diagnosis information from websites or medical institutes, labeling vast amounts of skin disease data requires expertise knowledge and can be really difficult and expensive in terms of both time and money. As is known that training a deep neural network requires a large amount of labeled data. Overfitting tends to occur when only small dataset is available. Therefore, larger datasets with labeled information are in demand to train an effective deep neural network for skin disease diagnosis. However, considering the practical challenges in developing a large dataset, it is also imperative to simultaneously develop approaches that exploit deep learning with less labeled data for skin disease diagnosis.
%Moreover, most existing skin disease datasets only contain data of the most common skin diseases (e.g., melanoma, seborrheic keratosis and benign nevi). As a result, deep learning methods trained with these datasets are incapable of diagnosing other kind of skin diseases that are not included in these datasets.
%\begin{itemize}\item[$\bullet$]

\subsubsection{Imbalanced skin disease datasets}
One common problem occurred in skin disease diagnosis tasks is the imbalance of samples in skin disease datasets. Actually, many datasets contain significant disproportions in the number of data points among different skin classes and are heavily dominated by data of the benign lesions. For example, one skin disease dataset may contain a large number of negative samples but only limited positive samples. Training deep learning models with imbalanced data may result in biased results, despite employing training tricks such as penalization of false negative cases found in a minor skin lesion class with weighted loss function. In light of the low frequency of occurrences of certain positive samples in skin diseases, obtaining a balanced dataset from the available original data can be as hard as developing a large-scale dataset.

\subsubsection{Noisy data obtained from heterogeneous sources}

Dermoscopy images of most existing skin disease datasets are obtained with high-resolution DSLR cameras in an optimal environment of lighting and distance of capture. Deep learning algorithms trained on these high-quality skin disease datasets are capable of achieving excellent diagnostic performance. However, when tested with images captured with low-resolution cameras (e.g., cameras of smart phones) in different lighting conditions and distances, the same model may be hard to achieve the same performance. Actually, deep learning algorithms are found to be highly sensitive to images captured by different equipments. In addition, self-captured images are often of inferior-quality with much noise. Therefore, noisy data obtained from heterogeneous sources bring challenges to skin disease diagnosis with deep learning.

\subsubsection{Lack of diversity among cases in existing skin disease datasets}

Most cases in existing skin disease datasets are fair-skinned individuals rather than dark-skinned ones. Though the incident rate of skin cancer is relatively higher among fair-skinned population than that of the dark-skinned population, people with dark skin can also suffer from skin cancer and are usually diagnosed in later stage~\cite{hu2006comparison}. Deep learning algorithms trained with skin disease data of fair-skinned population may fail to diagnose for the people with dark skin~\cite{marcus2019rebooting}. Another problem with existing skin disease datasets is that only categorizes of high incident rate (e.g., BCC, SCC and melanoma) are included and other (e.g., Merkel cell carcinoma (MCC), appendageal carcinomas, cutaneous lymphoma, sarcoma, kaposi sarcoma, and cutaneous secondaries) are ignored. Consequently, if deep learning algorithms are trained on datasets that do not contain data captured from dark-skinned population and have not adequate cases of rare skin diseases, misdiagnosis on data with these skin conditions may occur with a high probability.
Therefore, developing skin disease datasets with high diversity is significant for constructing effective skin disease diagnosis systems.

\subsubsection{Missing of medical history and clinical meta-data of patients}

Besides performing visual inspection for a suspected skin lesion with the help of medical equipment (e.g., dermoscopy), clinicians also take the medical history, social habits and clinical meta-data of patients into account when making a diagnostic decision. Actually, it is of great importance to know the diagnostic meta-data, such as skin cancer history, age, sex, ethnicity, general anatomic site, size and structure of skin lesions of patients (sometimes related information of their families are also needed). It has been proved in the work~\cite{haenssle2018man} that the performance of the beginner or skilled dermatologists can be improved with additional clinical information. However, most existing works on skin disease diagnosis with deep learning merely considered skin images and ignored medical history and clinical information of patients. One possible factor leading to this situation is the missing of such information in the most publicly available skin disease datasets.

\subsubsection{Explainability of deep learning methods}
There has been much controversy about the topic of ``black box" of deep learning models. That is, people may not be possible to understand how the determination of output is made by deep neural networks. This opaqueness has led to demands for explainability before a deep learning algorithm can be applied to clinical diagnosis. Clinicians, scientists, patients, and regulators would all prefer having a simple explanation for how a neural network makes a decision about a particular case. In the example of predicting whether a patient has a disease, people would like to know what hidden factors the network is using. However, when a deep neural network is trained to make predictions on a large dataset, it typically uses its layers of learned, nonlinear features to model a huge number of complicated but weak regularities in the data. It is generally infeasible to interpret these features since their meaning depends on complex interactions with uninterpreted features in other layers. If the same network is refit to the same data but with changes in the initializations, there may be different features in the intermediate layers. This indicates that unlike models in which an expert specifies the hidden factors, a neural network has many different and equally good ways to model the same dataset. It is not trying to identify the ``correct" hidden factors, but merely use hidden factors to model the complicated relationship between the input variables and output variables. In the future, more efforts should be made to deal with the ``black box" phenomenon.

\subsubsection{Selection of deep neural networks for a specific skin disease diagnosis task}
As the literature presented in the previous sections showed that most existing skin disease diagnosis tasks typically employed the currently popular deep architectures for image segmentation or classification. Additionally, ensemble methods of combining two or more deep networks were also utilized to analyze skin images. However, few works have made it clear how to select an appropriate type of deep neural network for a specific skin disease diagnosis task. Therefore, it is necessary to investigate the characteristics of skin diseases and corresponding data, and then design deep networks with domain knowledge for the specific task. In this way, better performance can be achieved.

\subsection{What can we do next?}

With the increasing trend of applying deep learning methods to skin disease diagnosis recently, people are likely to witness a large number of works in this field in the near future. However, as discussed above, several challenges exist and need to be resolved in this field. To cope with the challenges and obtain satisfying performance for skin disease diagnosis, there are a few possible directions that we can explore. We draw insights from the literature in the field of skin disease diagnosis and other fields (e.g., computer vision and pattern recognition), and present possible guidelines and directions for future works in the following.

\subsubsection{Obtain massive labeled skin disease data}
To obtain excellent performance for skin disease diagnosis, deep neural networks commonly require large amounts of data for training. However, limited labeled skin disease data is common in practice. To deal with this problem, we can seek solutions from several aspects. On one hand, people may employ experienced clinicians to label skin disease data manually, though it would be expensive and time-consuming. One the other hand, automated or semi-automated data labeling tools, such as Fiji~\cite{schindelin2012fiji}, LabelMe~\cite{russell2008labelme} and Imagetagger~\cite{fiedler2018imagetagger}, can be utilized to label massive data efficiently. Moreover, existing publicly available skin datasets can be comprehensively integrated to form a large-scale skin image dataset, as ImageNet in the computer vision field, for testing deep learning algorithms. In addition, to cope with the issue caused by noisy data with heterogenous sources, color constancy algorithms, such as Shades of Gray, max-RGB, can be utilized to boost the performance of deep learning models~\cite{barata2014improving,hua2019effect}. These algorithms can be used as image preprocessing methods to normalize the lighting effect of dermoscopy images.

\subsubsection{Increase the diversity of clinical skin data}
From the previous section we can observe that only limited skin disorders were involved in most works on skin disease diagnosis with deep learning methods~\cite{liao2016deep,brinker2019convolutional,han2018classification,esteva2017dermatologist}. As a result, the trained algorithms can only decide whether a lesion is more likely a predefined type of skin disease, such as nevus or melanoma, without even determining any subtype of it. By contrast, an experienced pathologist can diagnose any given images of a broad spectrum of differential diagnoses and decide a skin lesion belonging to any possible subtype of a skin disease. A more powerful and reliable skin disease diagnosis system that can be adapted to analyze all kinds of skin lesions is in huge demand. Consequently, it is necessary to expand the existing skin image datasets to include other cutaneous tumors and normal skin types. Moreover, it is also imperative to include skin data captured from the dark-skinned population to improve the diversity of current skin datasets. In this way, deep learning models trained on these general and complex datasets can adapt to more general skin disease diagnosis tasks.

%Though many deep learning algorithms have achieved excellent performance on existing skin disease datasets. However, the effectiveness of the algorithms should be further evaluated on more complex datasets. Particularly, prospective studies implemented in clinical settings are necessary to confirm a clinical impact of deep neural networks in assisting skin disease diagnosis. Increasing the data diversity is beneficial to construct general and complex datasets. Consequently, deep learning algorithms can be effectively trained and fully evaluated on these datasets before applied to practical tasks.

\subsubsection{Include additional clinical information to assist skin disease diagnosis}
In most cases, only dermoscopy or histopathological images are input to deep learning models for skin disease diagnosis. However, in the clinical settings, accurate diagnosis also relies on the history of skin lesions, risk profile of individuals, and global assessment of the skin. Thus, dermatologists commonly incorporate additional clinical information to identify skin cancers. The authors in \cite{haenssle2018man} investigated the effect of including additional information and close-up images for skin disease diagnosis and found a great improvement in the performance. Therefore, additional clinical information can be incorporated into the model training and testing processes for skin disease diagnosis. Other existing medical record data, such as un-organized documents, can be processed with techniques including NLP, document analysis~\cite{xu2017quary} and data mining~\cite{wu2013data} and taken into account in the diagnosis process as well. Skin images and related medical documents can be combined together to construct multi-view paradigms for the diagnosis tasks. Multi-view models have proved their effectiveness in recent works and can be extended to the field of skin disease diagnosis. Besides, integrating human knowledge into existing deep learning algorithms is likely to further improve the diagnosis performance as well.

\subsubsection{Fuse handcrafted features with deep networks extracted features}
Handcrafted features are typically extracted with less powerful traditional machine learning models and can be obtained with relatively small labeled data and less computational cost. However, they sometimes can achieve excellent performance in certain skin disease diagnosis tasks. Though handcraft features commonly lack generalization properties and showed inferior performance compared with the features directly learned from massive data with deep neural networks, they can be served as a supplementary to deep features. For example, decorrelated color spaces can be investigated to analyze the impact of color spaces in border detection and use them to facilitate skin image processing~\cite{2019Improved}. Skin lesion elevation and evolution features and geometrical features provide important clues for diagnosing a skin disease. Combining these features with deep features can further enhance the performance of current deep learning methods. Particularly, it would be promising if one could find a way to integrate the handcrafted feature extracting process with the learning process of deep networks. Through fusing handcrafted features with deep features, we may not only reduce the requirement of large amounts of labeled data to train a deep network, but also achieve better performance. Additionally, we need to investigate the characteristics of skin diseases first, and then design deep networks with domain knowledge for the specific task. In this way, better performance can be expected.% Decorrelated color spaces of skin images can be explored to promote the performance. Image segmentation is a significant step for skin disease diagnosis. Decorrelated color spaces can be investigated to analyze the impact of such color spaces in border detection. Performance evaluation metrics, such as Jaccard index and dice coefficients, can also be incorporated into the loss function of the deep neural networks so as to further improve the performance. %Deep CNNs pretrained on ImageNet may yield different results when compared with a counterpart that was trained on other datasets~\cite{menegola2017knowledge}. Investigating the influence of other datasets is beneficial for skin disease diagnosis.

\subsubsection{Employ GANs to synthesize additional data for training deep networks}
GANs~\cite{goodfellow2014generative} are attracting lots of attention from the computer vision community due to their ability to generate realistic synthetic images for various tasks. Then these images can be utilized as additional labeled data to train deep learning models. In this way, models commonly show superior performance compared with the situation where models are trained with limited original data. This property of GANs can be of great help for skin disease diagnosis when large-scale labeled datasets are unavailable. Actually, there have been a few works in the literature applying GANs to skin disease diagnosis~\cite{udrea2017generative,izadi2018generative,yi2018unsupervised,2020AQin}. However, it should be very careful when exploiting GANs for medical applications. As we know, GANs are trying to mimic the realistic images instead of learning the original distribution of images. Thus, images generated with GANs can greatly differ from the original ones. In light of this, it is feasible to train a deep learning model with images generated by GANs at the beginning and then fine-tune the final model with only the original images.

\subsubsection{Exploit transfer learning and domain adaptation for skin disease diagnosis}
Transfer learning~\cite{shin2016deep,tan2018survey} and domain adaptation~\cite{tzeng2017adversarial,you2019universal,cao2019learning} have been exploited to deal with the issues caused by lack of large-scale labeled data. As presented above, there have been many works utilizing transfer learning or domain adaptation techniques to improve the performance of deep learning models in skin disease diagnosis tasks~\cite{gu2019progressive,polevaya2019skin,jaworek2019melanoma,soudani2019image}. One way to implement transfer learning is to utilize existing pretrained deep learning models to extract semantic features and perform further learning based on these features~\cite{akhtar2017joint,bar2015deep,lopes2017pre}. For instance, Akhtar et al.~\cite{akhtar2017joint} utilized deep models to extract features and these features were further used to learn higher level features with dictionary learning. Another way to implement transfer learning is freezing part of a deep network and training the remainder. It is known that the initial layers of a deep network learn similar filters from diverse images. Therefore, one can directly borrow the values of parameters corresponding to initial layers from a network trained in similar tasks and freeze these layers. Then the remainder of the network is trained as normal with limited labeled data. In addition, we can take advantage of recent development~\cite{Haritha2020An} of transfer learning in the other fields (e.g., computer vision) to facilitate the success of deep learning in skin disease diagnosis tasks.% For example, Li et al.~\cite{li2019delta} proposed a novel regularized deep transfer learning framework using feature map with attention for CNNs. Instead of constraining the weights of networks, the proposed framework regularized the difference between feature maps generated by the convolution layers of the source and target networks with attentions.

\subsubsection{Develop semi-supervised deep learning methods for skin disease diagnosis}
It is known that large amounts of labeled data is required to train a deep learning model. However, collecting massive labeled skin data is expensive since expert knowledge is required and the labeling process is time-consuming. By contrast, it is much easier or cheaper to obtain large-scale unlabeled skin data. Semi-supervised learning~\cite{chapelle2009semi} aims to greatly alleviate the issues caused by lack of large-scale labeled data by allowing a model to leverage the available massive unlabeled data. Particularly, there have been a few works~\cite{masood2015self,yi2018unsupervised,He2020Semi} involving semi-supervised learning for skin disease diagnosis. Recently, semi-supervised deep learning attracts increasing attention in the field of computer vision and a few successful models have been proposed~\cite{berthelot2019mixmatch,Xiang2020FMixCutMatch,Dupre2020Improving}. Understanding these models and developing semi-supervised deep learning models specifically for skin disease diagnosis can be a promising direction.

\subsubsection{Explore the possibility of applying reinforce learning for skin disease diagnosis}
Reinforce learning (RL)~\cite{mnih2015human,jonsson2019deep} has achieved tremendous success in recent years, reaching human-level performance in several areas such as Atari video games~\cite{mnih2015human}, the ancient games of Go~\cite{silver2016mastering} and chess~\cite{silver2017mastering}. The success in part has been made possible by the powerful function approximation abilities of deep learning algorithms. Many medical decision problems are by nature sequential; therefore, RL can be employed to solve these problems. Particularly, there have been several works utilizing RL to solve medical image processing tasks and achieved promising results~\cite{ghesu2017multi,sahba2006reinforcement,netto2008application}. To the best of our knowledge, there have not works applying RL to skin disease diagnosis tasks so far. Therefore, RL can be a potential tool to solve skin disease diagnosis problems.

\subsubsection{Reasonable explanation for predictions produced by deep learning algorithms}

Explainability is one of the key factors that hinders the application of deep learning methods to clinical diagnosis scenarios. To assist diagnosis, people need reasonable explanation for the predictions produced by deep learning algorithms rather than just a confidence score of the skin diseases. One possible solution to this problem is to provide a reasonable explanation for the predictions according to the ABCDE criteria (asymmetry, border, color, diameter, and evolution) or $7$-point skin lesion malignancy checklist (pigment network, regression structures, pigmentation, vascular structures, streaks, dots and globules, and blue whitish veil)~\cite{goyal2019artificial}.

\section{Summary}

In this review, we present an overview on the advances in the field of skin disease diagnosis with deep learning. First, we briefly introduce the domain and technical aspects of skin disease. Second, skin image acquisition methods and publicly available datasets are presented. Third, the conception and popular architectures of deep learning and commonly used deep learning frameworks are introduced. Then, we introduce the performance evaluation metrics and review the applications of deep learning in skin disease diagnosis according to the specific tasks. Thereafter, we discuss the challenges remained in the area of skin disease diagnosis with deep learning and suggest possible future research directions. Finally, we summarize the whole article.

Compared with existing related literature reviews, this article provides a systematic survey of the field of skin disease diagnosis focusing on recent applications of deep learning. With this article, one could obtain an intuitive understanding of the essential concepts in the field of skin disease diagnosis with deep learning and challenges faced in this field as well. Moreover, several possible directions to deal with these challenges can be taken into consideration by ones who are willing to work further in this field in the future.

The potential benefits of automated diagnosis of skin diseases with deep learning are tremendous. However, accurate diagnosis increases the demand of reliable automated diagnosis process that can be utilized in the diagnostic process by experts and non-expert clinicians. From the review, we can observe that numerous deep learning systems have been proposed and achieved comparable or superior diagnosis performance on experimental skin disease datasets. However, we should be aware that a computer-aided skin disease diagnosis system should be critically tested before it is accepted for real-life clinical diagnosis tasks.

\section*{Acknowledgment}
This work was supported by the National Key Research and Development Program of China under Grant 2018YFC0910700 and 2019YFC0840706, the Clinical Medicine Plus X - Young Scholars Project, Peking University 71006Y2408 and the National Natural Science Foundation of China 11701018.

\section*{References}

\bibliography{mybibfile}

\end{document}